\documentclass[12pt]{article}
\usepackage{graphicx}
\usepackage{epsfig}
\usepackage{overcite}
\usepackage{color}
\newcommand{\frn}{$f(r_n)$}
\newcommand{\rn}{$\overline{r}_n$}

\begin{document}

\begin{center}

\noindent {\LARGE {\bf A study of the size dependence of self diffusivity across the solid-liquid transition}}

\noindent{\bf Manju Sharma$^1$ and S. Yashonath$^{1,2,\dagger}$}\\

\baselineskip=12pt
\noindent {\it $^{1}$ Solid Sate and Structural Chemistry Unit}\\
\noindent {\it $^{2}$ Center for Condensed Matter Theory}\\
\noindent {\it Indian Institute of Science, Bangalore, India - 560 012}\\
\end{center}

\footnotetext {$\dagger$  Also at Jawaharlal Nehru Centre for Advanced
Scientific Research, Jakkur, Bangalore }

\vspace*{1.0cm}
\begin{center}
{\large\bf Abstract}
\end{center}
\noindent
\baselineskip=12pt

The present study investigates the effect of temperature-induced change of phase from face centred cubic
solid to liquid on the size dependence of self diffusivity of solutes through detailed 
molecular dynamics simulations on binary mixtures consisting of a larger solvent
and a smaller solute interacting via Lennard-Jones potential. The effect of change 
in density during the solid-liquid 
transition as well as in the solid and the liquid phase itself have been
studied through two sets of simulations one at fixed density (NVE simulations) 
and another at variable density (isothermal-isobaric ensemble). 
A size dependent diffusivity
maximum is seen in both solid and liquid phases at most densities except at low liquid densities.
Two distinct regimes may be identified : linear and anomalous regimes.
It is seen that the effect of solid to liquid transition is to shift the diffusivity maximum
to smaller sizes of the solutes and decrease the height of the maximum. 
This is attributed to the predominant influence of 
disorder (static and dynamic) and relatively weaker influence of density. The latter shifts
the diffusivity maximum towards larger solute size while the former does exactly the opposite.
Interestingly, we find that lower densities lead to weaker diffusivity maximum.
We report for the first time for systems with large solute-solvent interaction  
strength, $\epsilon_{uv}$, multiple diffusivity maximum is seen as a function of the solute size.
The origin of this is not clear. 
A remarkable difference in the ballistic-diffusive transition is seen between solutes
from the linear regime and anomalous regime. It is found that this transition is sharp for
solutes from anomalous regime.  Negative exponent $\alpha$ in dependence of 
mean square displacement on time $u^2(t)\ \sim\ t^\alpha$ is seen for linear regime
solute in solid phase during ballistic-diffusive transition which is completely absent 
for a solute from anomalous regime. This provides interesting insight into motion of anomalous regime
solutes. Apart from these, we report several properties which are different for 
the linear and anomalous regime solutes.

\baselineskip=22pt

\section {Introduction}

We have studied in previous chapters, size dependent diffusivity in close-packed 
body centred cubic solid phase and also liquid phase leading to breakdown in
Stokes-Einstien relationship between self diffusivity and solute diameter.
Several aspects of size dependence of diffusivity has been studied in these systems.
However, it is not clear how size dependent self diffusivity is altered across
the solid-liquid phase transition. Several question remain unanswered that pertain
to size dependent self diffusivity. Does solute diameter corresponding to diffusivity maximum
shift to larger size with melting since liquids are at a lower density as compared to
solids ? Can we isolate effects due to disorder and effects due to change in density ?

In this work, we report a study of size dependent diffusivity maximum across
the solid-liquid transition at (a) fixed density and (b) variable density. The
solid-liquid transition itself has been induced by increase in temperature rather
than changes in pressure. We locate the solid-liquid transition temperature
precisely for both fixed density and variable density simulations.

\section{Methods}

\subsection { Model and intermolecular potential}

Two series of runs have been made at a range of temperatures, one at fixed density referred to as FD and
another in which density was allowed to vary (variable density or VD) with temperature. 
The system being studied is a
two component system, consisting of a larger-sized solvent particles and a relatively smaller-sized
solute particles. Interactions between them is via simple (6-12) Lennard-Jones potential.
The solute-solvent Lennard-Jones interaction parameters have been derived from :
$\sigma_{uv}$=$\sigma_{uu}$+0.7\AA.\cite{Parr_rahman, Vash_rahman} The range of $\sigma$ and 
the value of $\epsilon$ chosen for solute-solute and solute-solvent and solvent-solvent 
interactions are listed in Table \ref{potn}. 
The total interaction energy of the system is a sum of solvent-solvent, $U_{vv}$, 
solute-solvent, $U_{uv}$ and solute-solute, $U_{uu}$ interaction energy.

\begin{equation}
U_{tot} = U_{vv} + U_{uv} + U_{uu}
\end{equation}

Initial configuration is face-centred cubic lattice of solvent with the solute occupying the
octahedral voids with this lattice. The simulations are then performed at 70K. For runs at
higher temperatures, the final configuration from the previous lower temperature has been
taken as the starting configuration.

\begin{table}
\caption {Lennard-Jones interaction parameters for the solute and solvent atoms.}
\begin{center}
\begin{tabular}{{c}{c}{c}}\hline
{type of interaction}&$\sigma$, \AA\ &$\epsilon$, kJ/mol \\\hline\hline
vv&4.5&1.2\\
uv&1.1-2.7&4.5\\
uu&0.4-2.0&0.4\\ \hline
\end{tabular}
\label{potn}
\end{center}
\end{table}

For variable density runs, we employed NPT ensemble at the desired temperature to
obtain an estimate of the correct density. We then made runs in the microcanonical
ensemble at this density. This ensures that the dynamics are computed in the microcanonical
ensemble without use of NPT ensemble.

\section {Computational Details}

We have chosen to study the system at 70, 120, 160 and 250K which includes
the solid-liquid transition. These are the temperatures for both fixed density as well as the 
variable density so that the different properties can be compared. 
The system consists of 1372 solvent and 147 solute species.
The solvent and solute masses are 40 and 20amu respectively. 
All the simulation runs have been performed using DLPOLY.\cite{dlpoly} 
The reduced density, $\rho^*$ for fixed density(FD) runs are 0.933 irrespective of temperature. 
For variable density NPT runs as well as the fixed density runs, the initial configuration for runs at
all temperatures is the face-centred cubic arrangement. The initial simulation cell length of this 
corresponded to a reduced density of $\rho^*$ = 0.933 which changed to values appropriate to the new
temperature. All variable density runs were made at a pressure of 0.4kbar. This pressure ensures that
the system is in the liquid phase at 160K similar to the fixed density run where the solid melts
at 159K. All NPT runs have been made with Berendsen method using $\tau_T$
and $\tau_P$ equal to 0.1 and 2.0ps respectively. 

In all NPT runs, a timestep of 2fs for 250K and 5fs for other temperatures was used with an 
equilibration of 500ps and a production run of 40ps during which the average density was 
computed for use in NVE runs. NVE simulations were made with Verlet leapfrog algorithm 
with an energy conservation of total energy of better than 1 in 10$^{-5}$. An equilibration 
period of 1ns is followed by a production run of 2ns for computing properties. 
The position coordinates, velocities and forces on solutes have been accumulated every 
250fs during the production run. The resulting cell dimension of the simulation cube are :
for FD runs : 51.1725\AA($\rho^*$ = 0.933), for VD runs : 48.7327\AA(1.0803, VD(70K)), 
49.7153\AA(1.0175, VD(120K)), 52.819\AA(0.8484, VD(160K)) and 
56.37\AA(0.6979, VD(250K)). 

A few runs have been made on a smaller system of 500 solvents and 50 solutes at 
$\rho^*$ = 0.933 and 70K at different values of $\epsilon_{uv}$ (Figure 7). 
Here, equilibration is for 1ns while production is for 500ps.

\section {Results and Discussion}

The reduced densities, $\rho^*$ of the systems in variable
density(VD) case, obtained from the simulations in isothermal-isobaric 
ensemble at different temperatures are reported in Table \ref{dens}. 
The system at temperatures 70K and 120K are at 1.0803 and 1.0175 density which are 
higher than the density of the FD run. At 160K and 250K the densities in VD
runs are 0.8484 and 0.6979, lower than the density of the  
fixed density(FD, $\rho^*$=0.933) simulation.

\begin{table}
\caption {The reduced densities of systems belonging to variable(VD) 
density case. The densities are obtained from simulations in 
isothermal-isobaric ensemble at 0.4kbar and temperatures,
70, 120, 160 and 250K.}

\begin{center}
\begin{tabular}{{c}{c}}\hline
T, K& $\rho^*$\\\hline\hline
70 &1.0803\\
120&1.0175\\
160&0.8484\\
250&0.6979\\
\end{tabular}
\label{dens}
\end{center}
\end{table}

\begin{figure}
\begin{center}
{\includegraphics*[width=8cm]{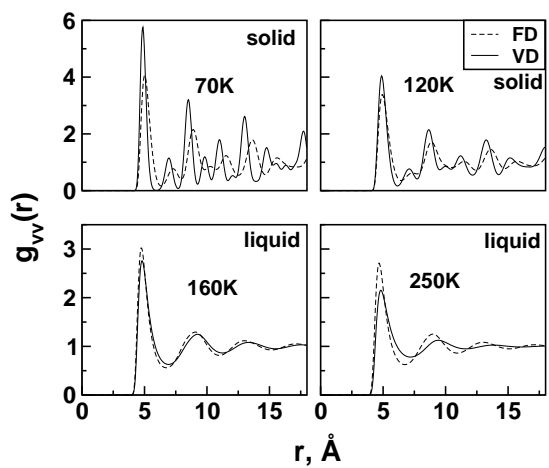}}
\caption{Solvent-solvent radial distribution plot in the solid and liquid phase for
the fixed(FD) and variable(VD) density systems.} 
\label{rdf}
\end{center}
\end{figure}

Figure \ref{rdf} shows the radial distribution(RDF) of solvent atoms at four temperatures
for both VD and FD simulations. The RDF shows several well defined peaks at 70K. 
The positions and intensities of the peaks correspond to the face-centred cubic arrangement. 
Thus, the solvent is in solid phase at 70K (for both VD and FD simulations).
The peaks are more sharp for the 
variable(VD, $\rho$*=1.0803) density system which is at higher density 
than the fixed (FD, $\rho$*=0.933) density system. With increase in 
temperature to 120K, RDF peaks are broader as compared to 70K and 
the system is still in the solid phase. Thus, both at 70 and 120K, the system
is in the solid phase and the VD simulations yield well ordered solid as compared to
FD simulations being at higher density. 

At 160K the system is in liquid phase as shown in the solvent-solvent radial 
distribution. The RDF peaks are broader in the variable(VD, $\rho$*=0.8484) 
density system as the reduced density is less than the fixed(FD, $\rho$*=0.933) 
density system, with the VD system being more disordered. Both FD and VD systems are 
highly disordered liquids by 250K. The variable (VD, $\rho$*=0.6979) density system at 
250K is the most disordered system among all the systems.

\begin{figure}
\begin{center}
{\includegraphics*[width=6cm]{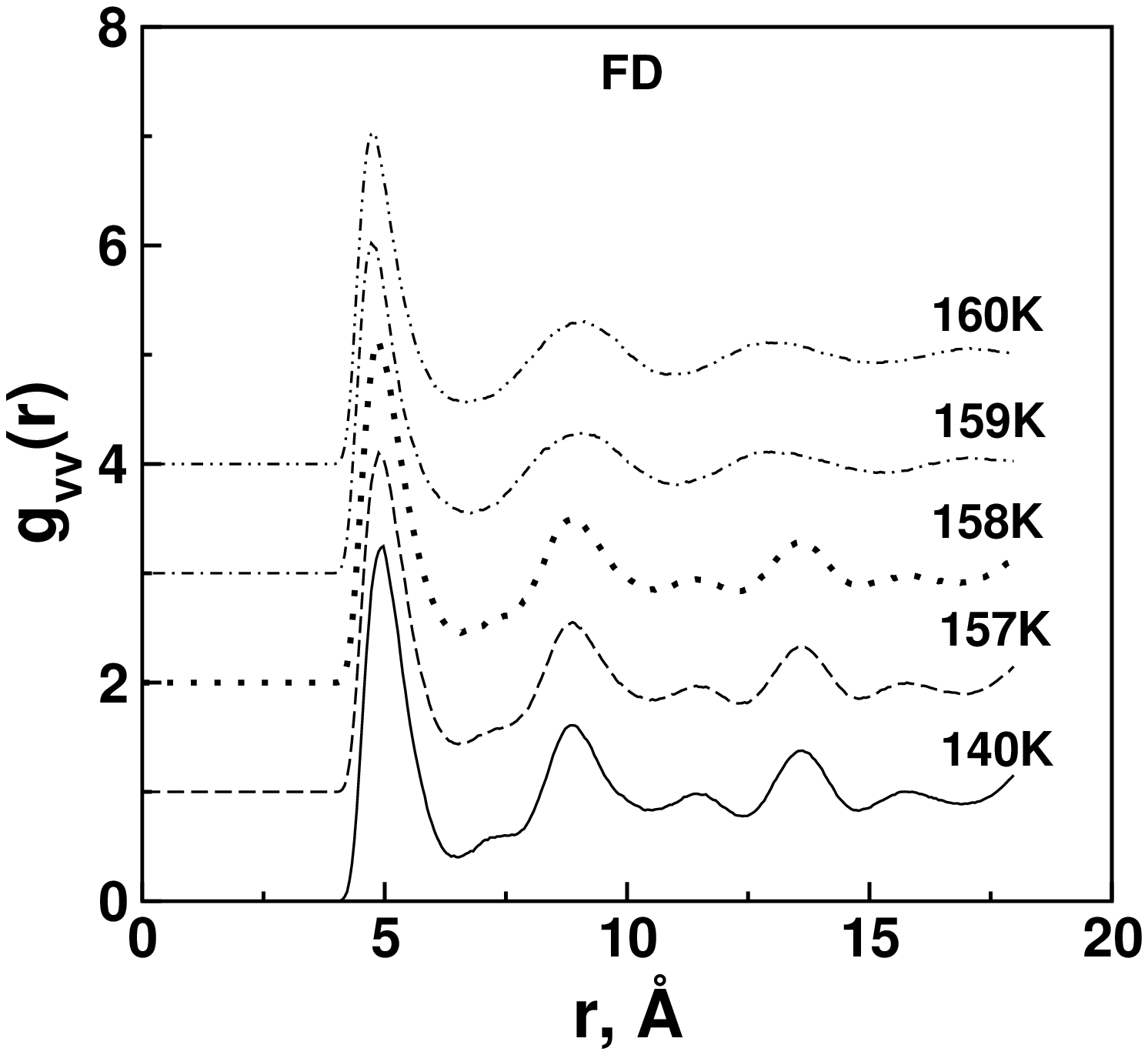}}\hspace*{0.5cm}
{\includegraphics*[width=6cm]{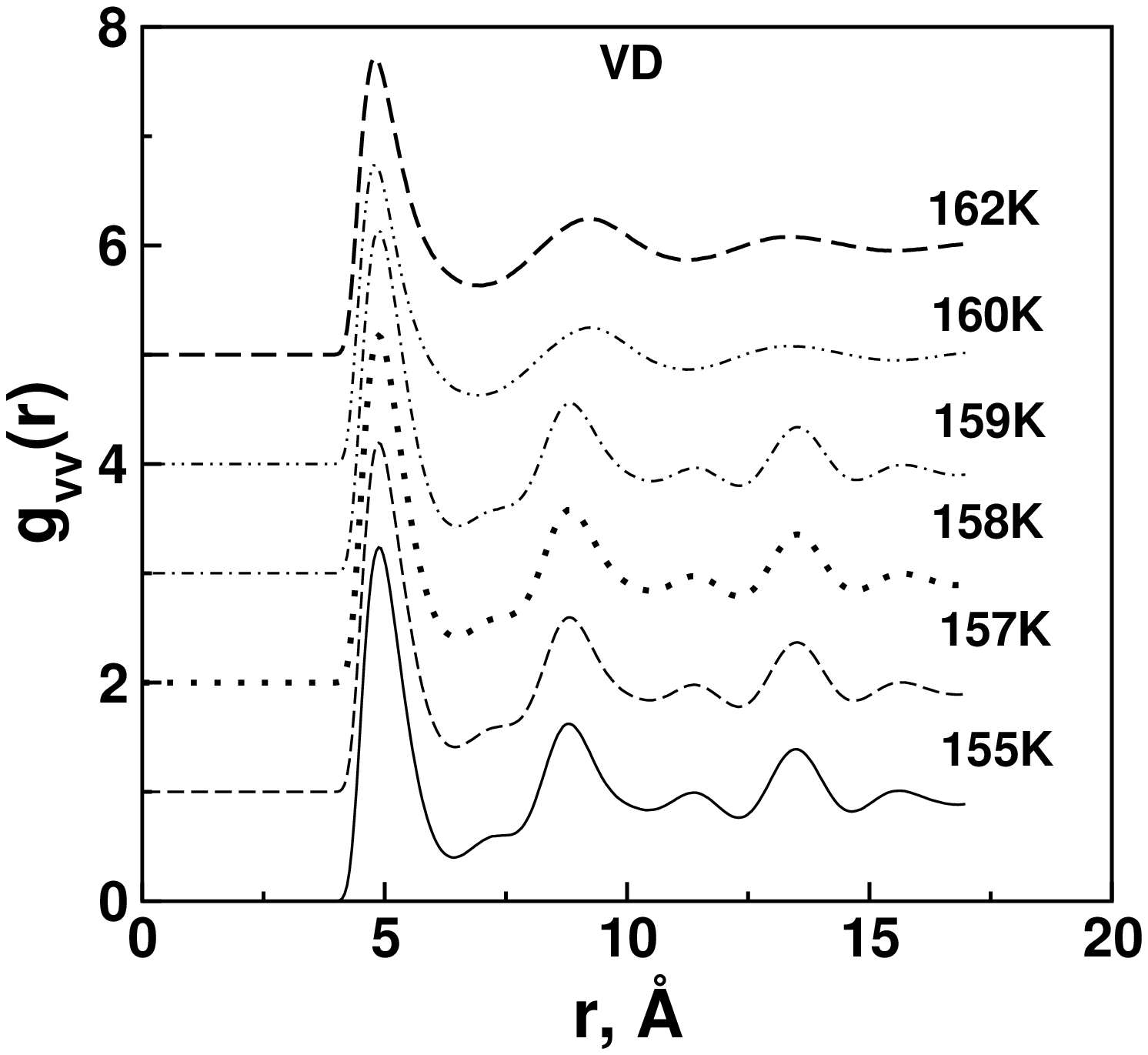}}
\caption{The transition temperatures as the system shifts from solid 
to liquid phase obtained from the solvent-solvent radial distribution at
different temperatures for the fixed density(159K) and varied(160K) density simulations.} 
\label{T_trans}
\end{center}
\end{figure}

The solvent system was simulated at different temperatures to obtain the
transition temperature from solid to liquid phase. Figure \ref{T_trans}
shows the solvent-solvent radial distribution at different temperatures. The
system melt at 159K as can be seen from the peaks of the radial distribution
plot in the case of FD simulations. The pressure was adjusted in the case of
VD simulations so as to obtain a transition temperature close to the transition
temperature for the FD simulations. We found this to be the case for a pressure
of 0.4kbar. Hence all simulations have been performed at 0.4kbar in case of
VD simulations. The transition temperature for VD simulations is 160K.

\begin{figure}
\begin{center}
{\includegraphics*[width=8cm]{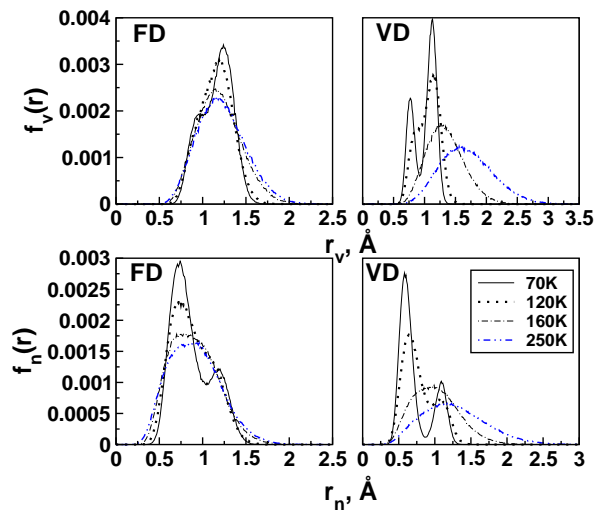}}
\caption{The normalized neck distribution of solvent atoms calculated using Voronoi tesselation 
as a function of phase and density.}
\label{neck}
\end{center}
\end{figure}

Further to understand the effect of change of phase and density 
on solvent network we have obtained the void and neck distribution of the solvent 
system in all the cases using Voronoi tesselation. \cite{Tanemura}
The neck is the interconnecting, usually narrower, region between any two voids. The neck size 
plays an important role in the size dependence of diffusion coefficient.\cite{pradipliq,
fea_art} Figure \ref{neck} reports both the void and the neck distribution of the solvent atoms. 
These have been normalized by dividing by the number of solvent particles and the number of
MD steps.
The neck distribution is bimodal at 70K with well defined distributions
for the variable($\rho$*=1.0803) density system. For the fixed density
system at $\rho$*=0.933 the distributions are less well defined as compared to VD.
With increase in temperature to 120K, there is a single distribution for FD but 
for VD system a small shoulder still persists.  
The fixed(FD, $\rho$*=0.933) system density is at lower density than the 
variable(VD, $\rho$*=1.0175) density and thus higher disorder in FD system
leads to complete merging of the second maximum with the first. At 160 and 250K,
when system exists in liquid phase, the maximum in the neck 
distribution shifts to higher sizes. Also the width of neck
distribution increases with increase in the temperature as a result of increased disorder.
As we shall see, these changes have profound impact on the size-depedent diffusivity maximum.
The void distribution exhibits similar changes.  

The average neck radii, $\overline{r}_n$ in all the systems are calculated 
from the neck distribution data using the Eq. \ref{av_neck}. 

\begin{equation}
\overline{r}_n = \frac{\Sigma f_n(r)r }{\Sigma f_n(r)}
\label{av_neck}
\end{equation}

For the solid phases where a bimodal distribution is seen, we have obtained two
average neck radii from the two distributions. The values of 
$\overline{r}_n$ for all the systems are reported in Table \ref{rn_Dmax}. 
The neck distribution and thus the average neck radii are significantly 
different for all the systems. In view of the previous result that the neck distribution
is responsible for the size-dependent diffusivity maximum, a pertinent question is 
how these changes affect the diffusivity maximum as the system changes 
from one phase to other?

\begin{table}
\caption {The average neck radii obtained from Voronoi tessellation
and the solute diameters for which diffusivity maximum is seen
at different temperatures and densities for fixed and variable density simulations in solid and liquid phases.}
\begin{center}
\begin{tabular}{{c}{c}{c}{c}{c}{c}}\hline
T, K & phase &  \multicolumn{2}{|c|} {fixed density}  & \multicolumn{2}{c|} {variable density} \\ \hline\hline
& & $\overline{r}_n$&, \AA  $\sigma_{uu}^{max}$, \AA & $\overline{r}_n$, \AA & $\sigma_{uu}^{max}$ \AA\\ \hline
70 &solid& 0.76, 0.87 & 1.1, 1.5 & 0.60, 0.73 & 1.1, 1.4\\
120&solid& 0.80 & 1.1 & 0.68, 0.79 & 1.1, 1.4 \\
160&liquid& 0.90 & 1.1  & 1.01 & 1.0 \\
250&liquid& 0.91 & 1.0  & 1.27 & \\ \hline
\end{tabular}
\label{rn_Dmax}
\end{center}
\end{table}

\begin{figure}
\begin{center}
{\includegraphics*[width=6cm]{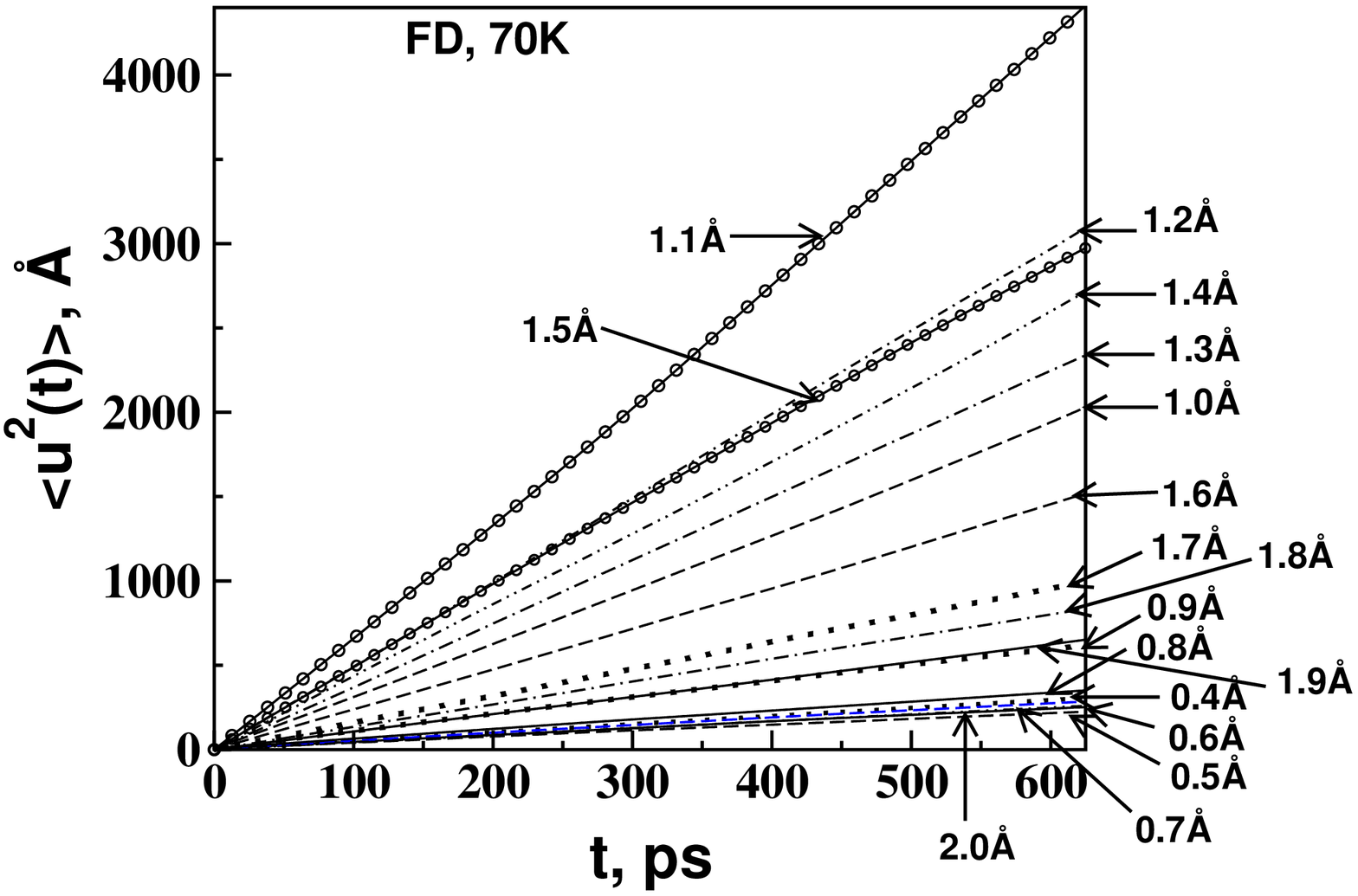}}\hspace*{0.5cm}
{\includegraphics*[width=6cm]{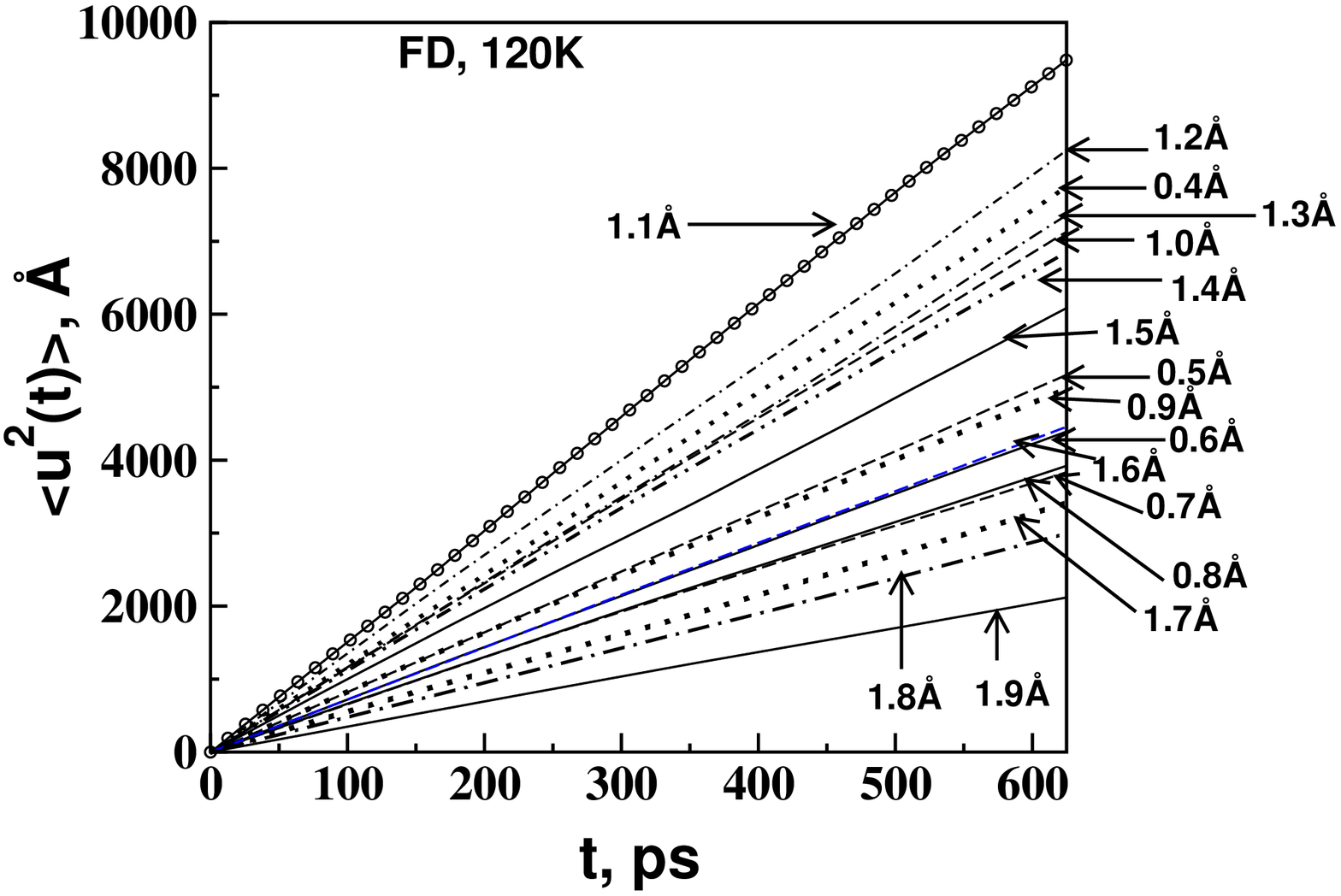}}\\
{\includegraphics*[width=6cm]{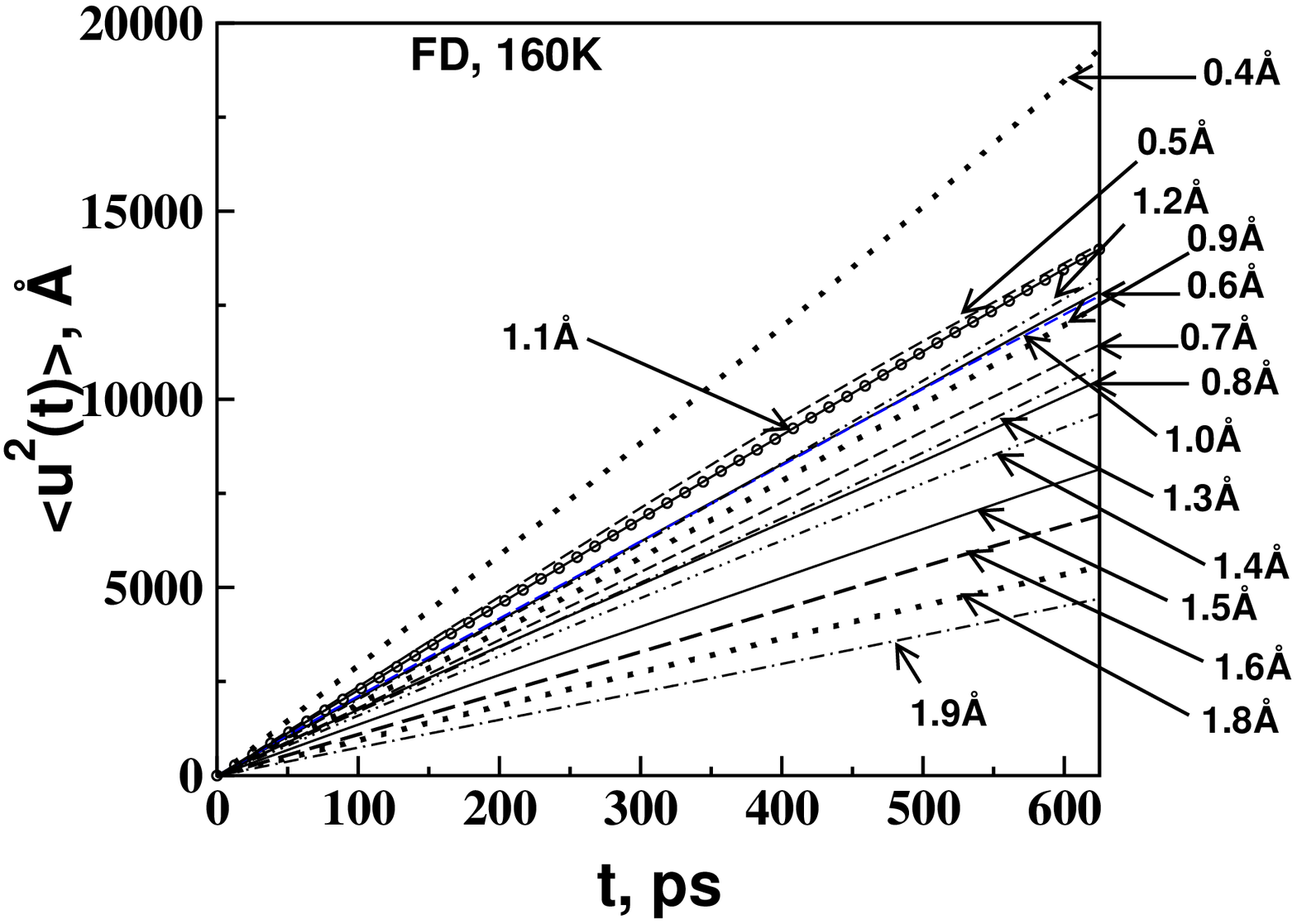}}\hspace*{0.5cm}
{\includegraphics*[width=6cm]{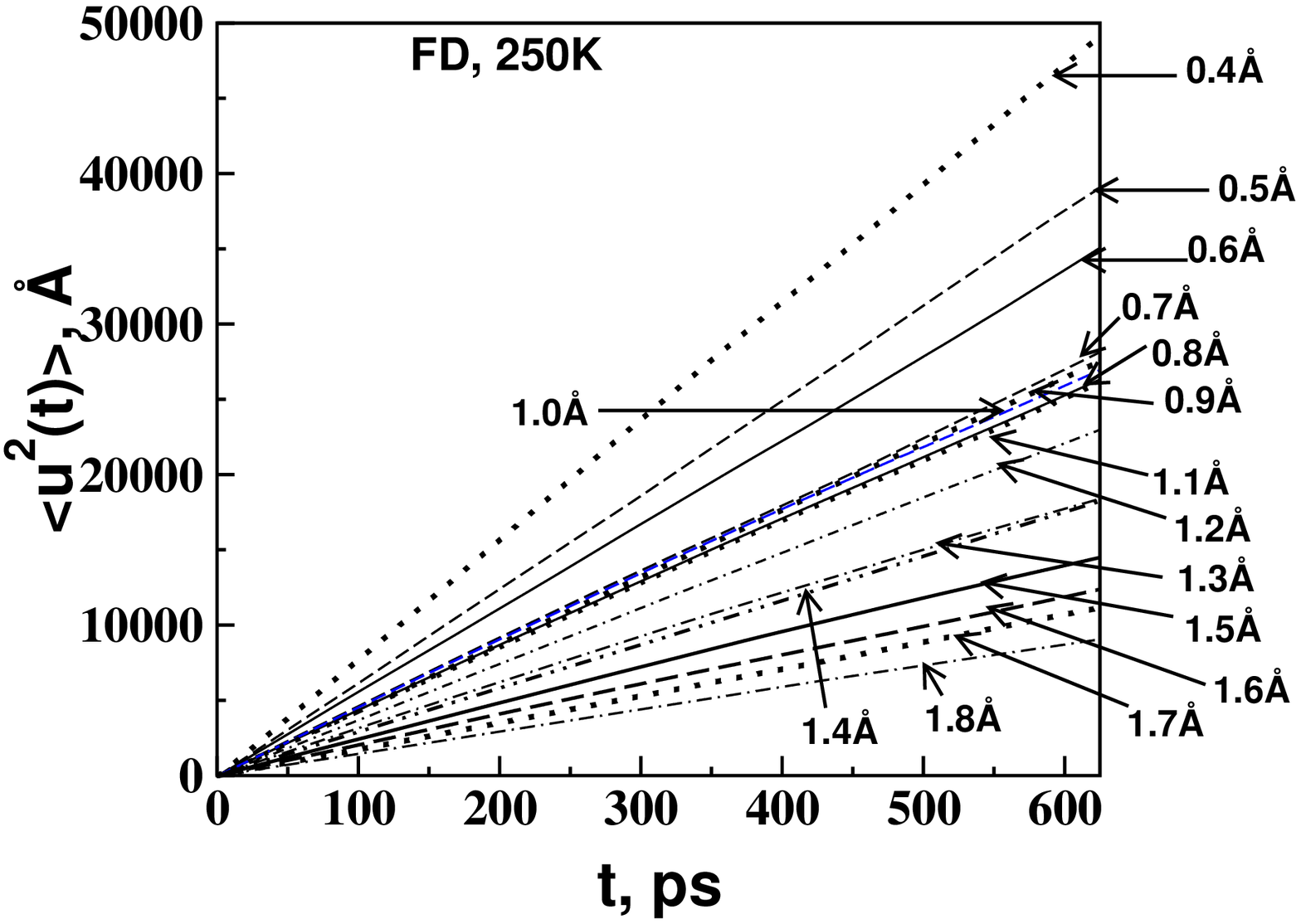}}\\
{\includegraphics*[width=6cm]{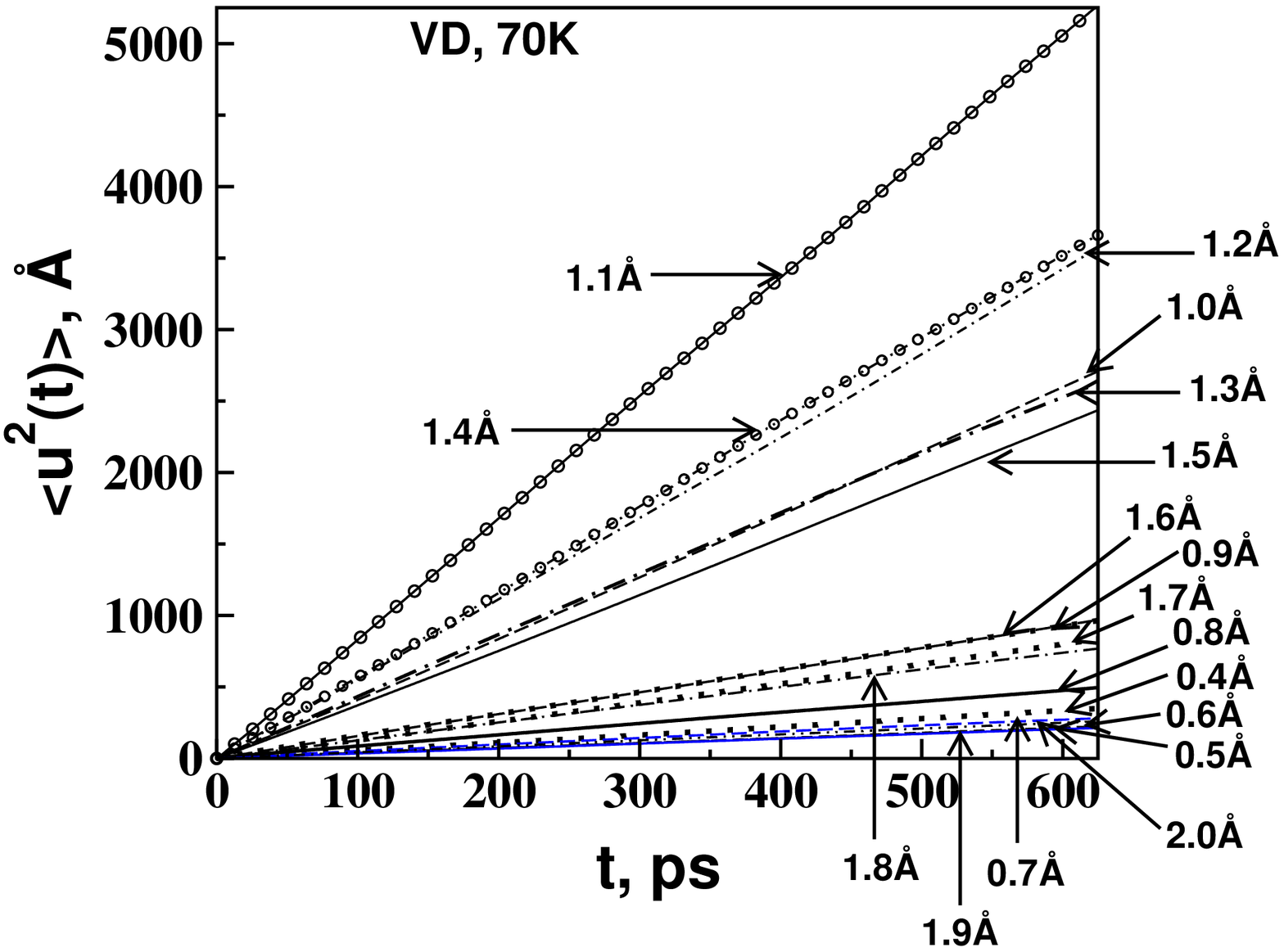}}\hspace*{0.5cm}
{\includegraphics*[width=6cm]{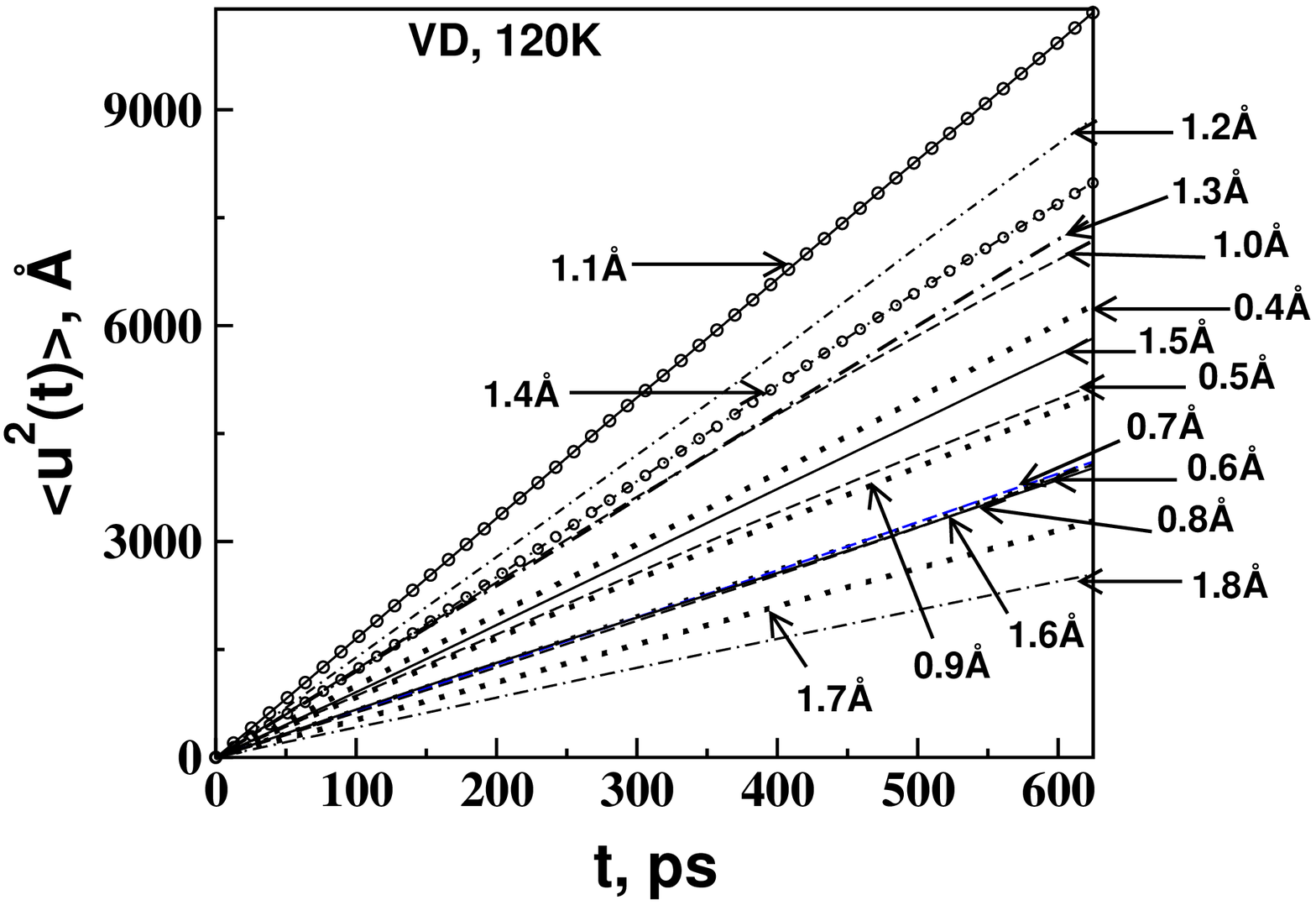}}\\
{\includegraphics*[width=6cm]{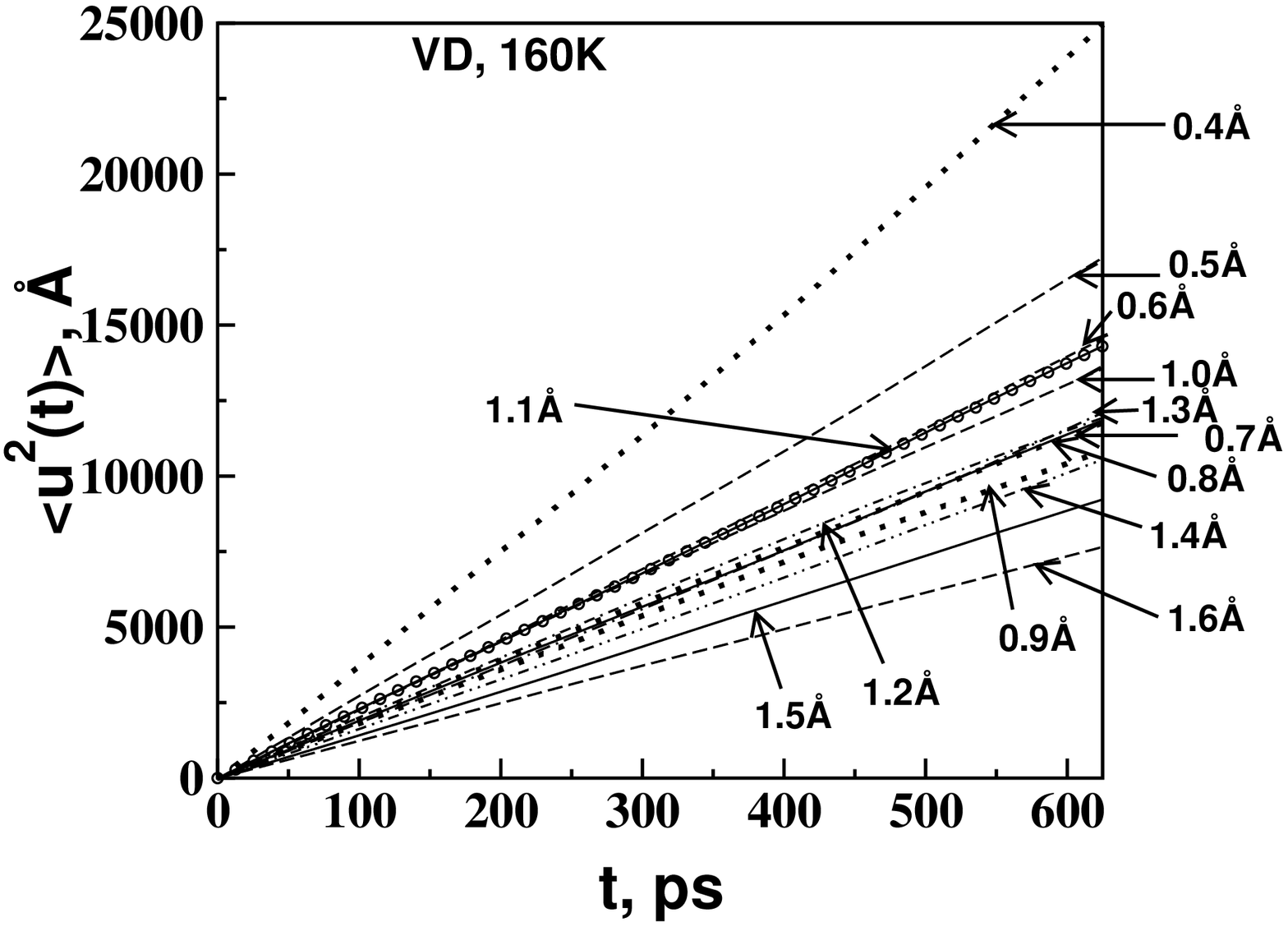}}\hspace*{0.5cm}
{\includegraphics*[width=6cm]{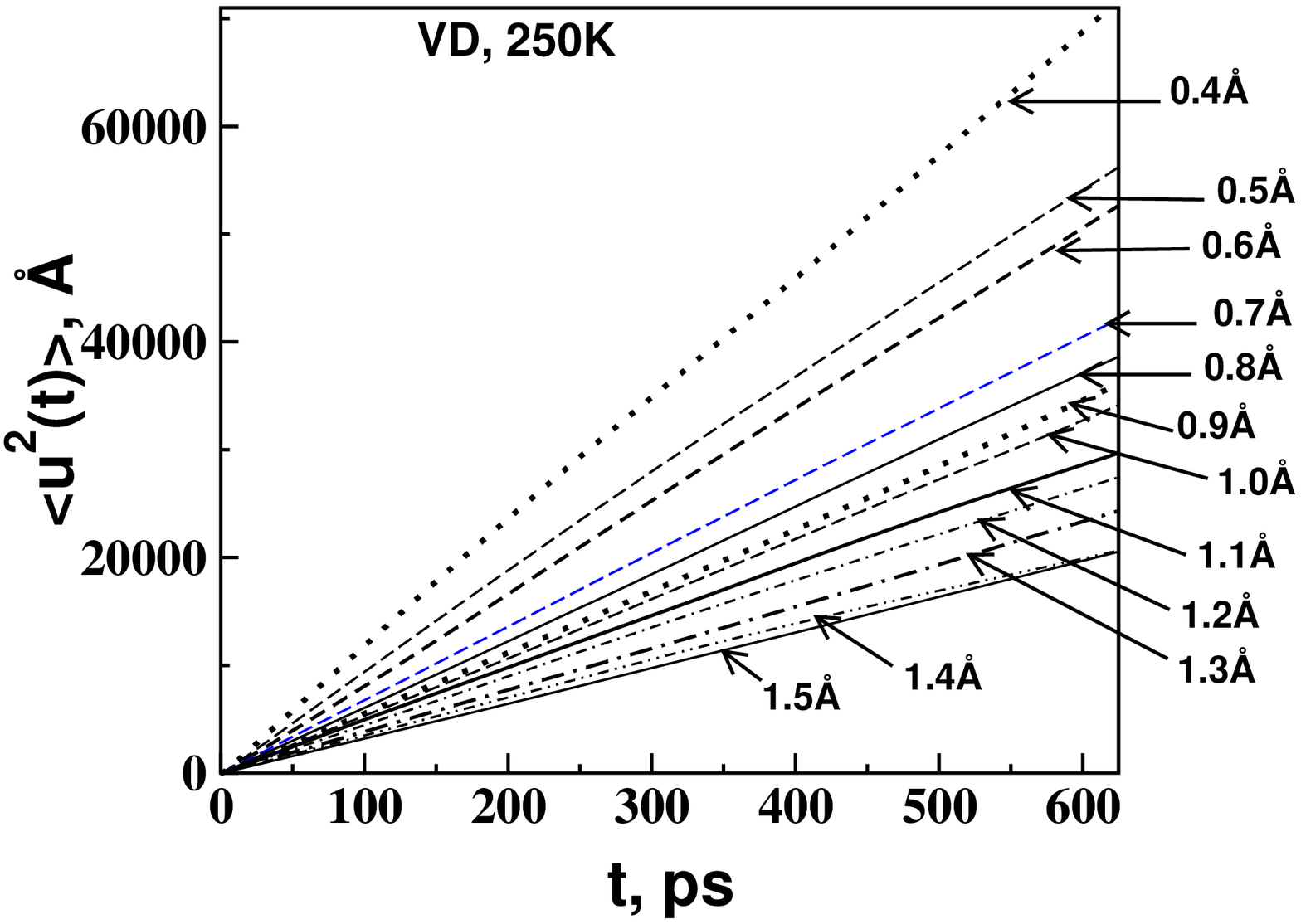}}

\caption{Time evolution of the mean square displacement of different sized solutes at four
different temperatures for VD and FD simulations in solid and liquid phases. }
\label{msd}
\end{center}
\end{figure}
\begin{figure}
\begin{center}
{\includegraphics*[width=8cm]{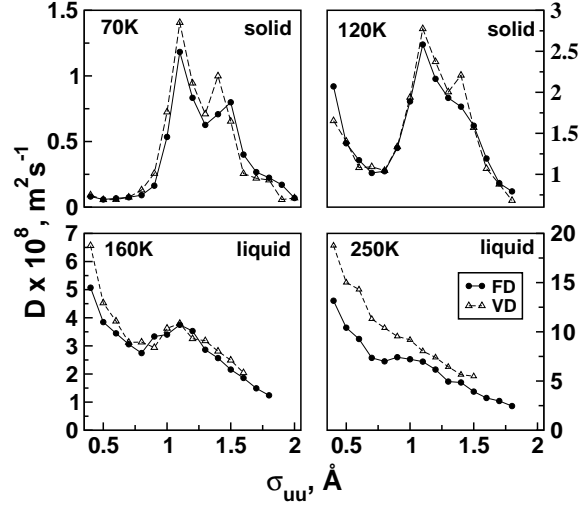}}
\caption{Impurity diffusivity as a function of size for the systems 
at fixed and variable densities.}
\label{D_sig}
\end{center}
\end{figure}

We have computed the mean square displacement (MSD), $u^2(t)$ for a range of solute diameter.
These are plotted in Figure \ref{msd}. Note that the msd curves are straight indicating 
that the statistics is good and therefore, the self diffusivity, $D$ are reliable.
The self diffusivities have been obtained from
the Einstein's relationship between the mean square displacement and diffusivity.
A plot of self diffusivity against solute diameter (the Lennard-Jones parameter $\sigma_{uv}$)
is shown in Figure \ref{D_sig}. Initially, for small sizes of the solute, $D$ decreases with
increase in solute diameter. Subsequently, an increase in the slope is seen for intermediate 
sizes of the solute. This is followed by a decrease in $D$ with increase in the size of the solute
at large sizes. Thus, a maximum is seen in $D$ for some intermediate size(s) of the solute.
We note that previously the small solute regime where $D$ decreases with increase in solute diameter
is called the linear regime (LR). The size from when $D$ increases with solute diameter and all sizes
greater than this size is referred to as anomalous regime (AR).\cite{yashosanti94b, pradipliq}
The maximum is seen when the diameter of the solute is comparable to the neck diameter 
in the distribution f($r_n$). Since neck diameter is not unique but a distribution of neck 
diameters exist, the diameter of the solute at which maximum in $D$ is seen, is influenced
by the distribution f($r_n$). 

\begin{figure}
\begin{center}
{\includegraphics*[width=8cm]{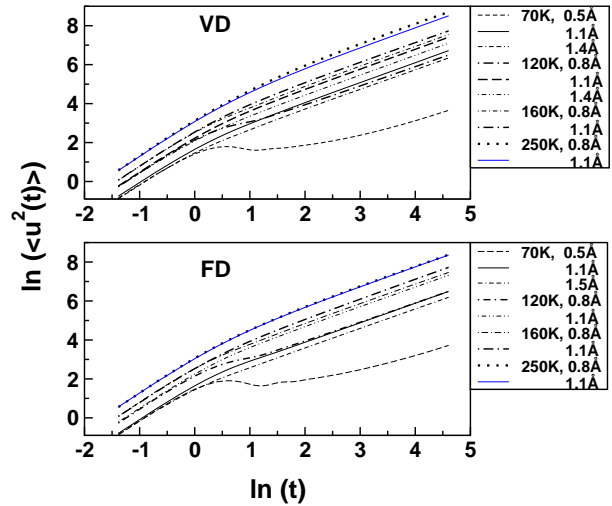}}\hspace*{0.5cm}
\caption {Logarithm of average mean square displacement as a function of logarithm
of time for solutes belonging to linear(LR) and anomalous(AR) regimes at different temperatures
for the fixed(FD) and varied(VD) density simulations. Note the ballistic to diffusive transition.}
\label{ln_msd}
\end{center}
\end{figure}

\begin{figure}
\begin{center}
{\includegraphics*[width=6cm]{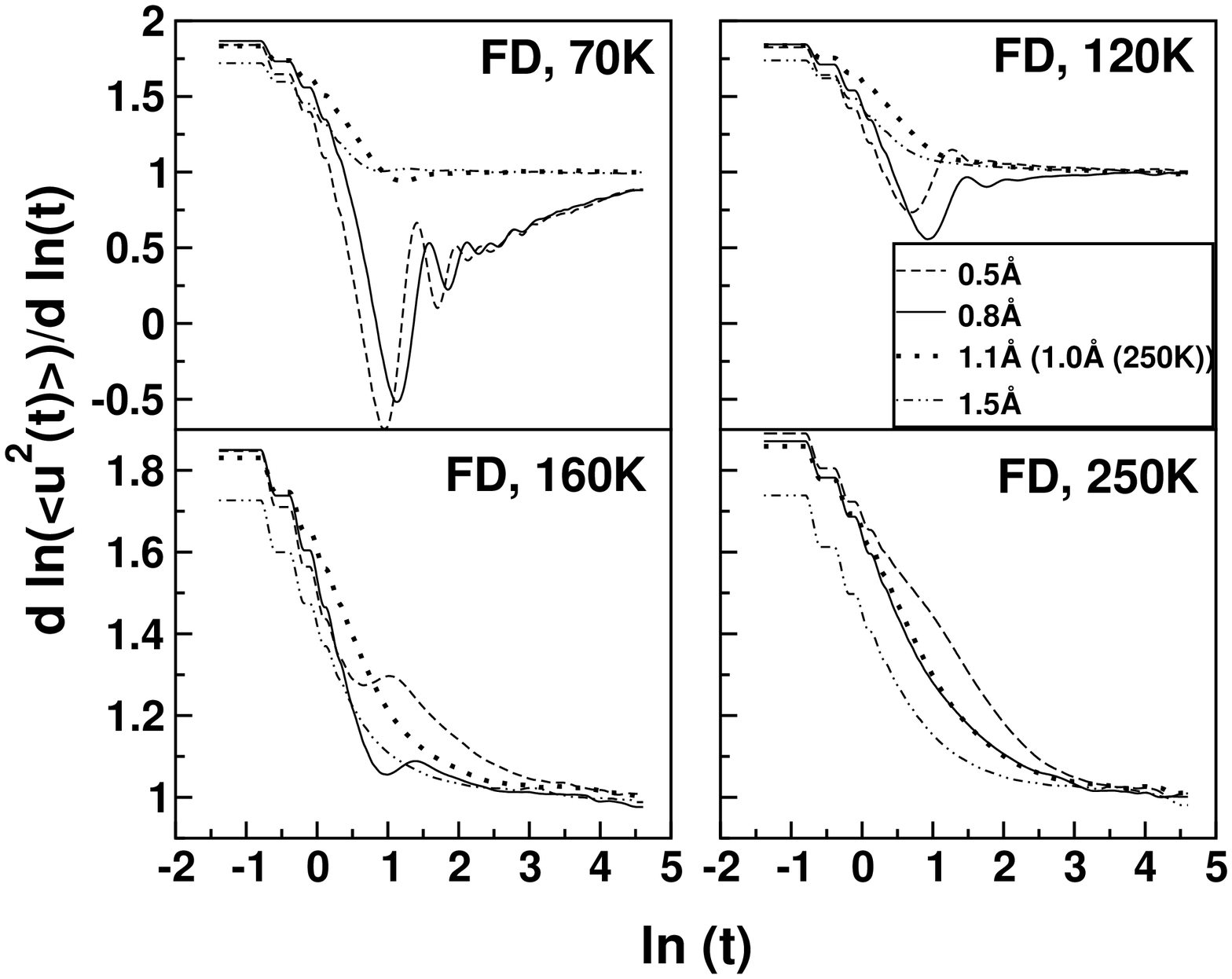}}\hspace*{0.5cm}
{\includegraphics*[width=6cm]{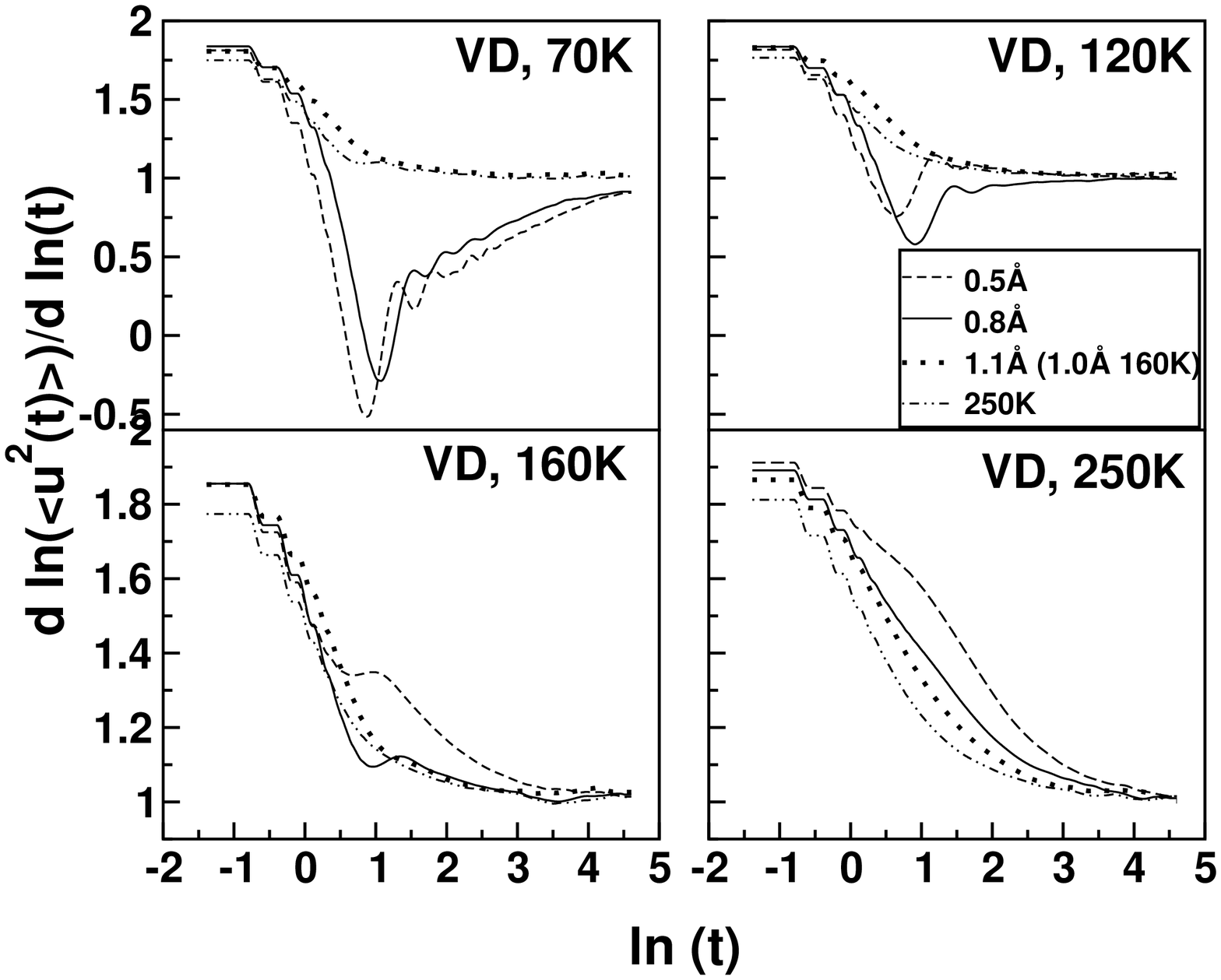}}\\
{\includegraphics*[width=6cm]{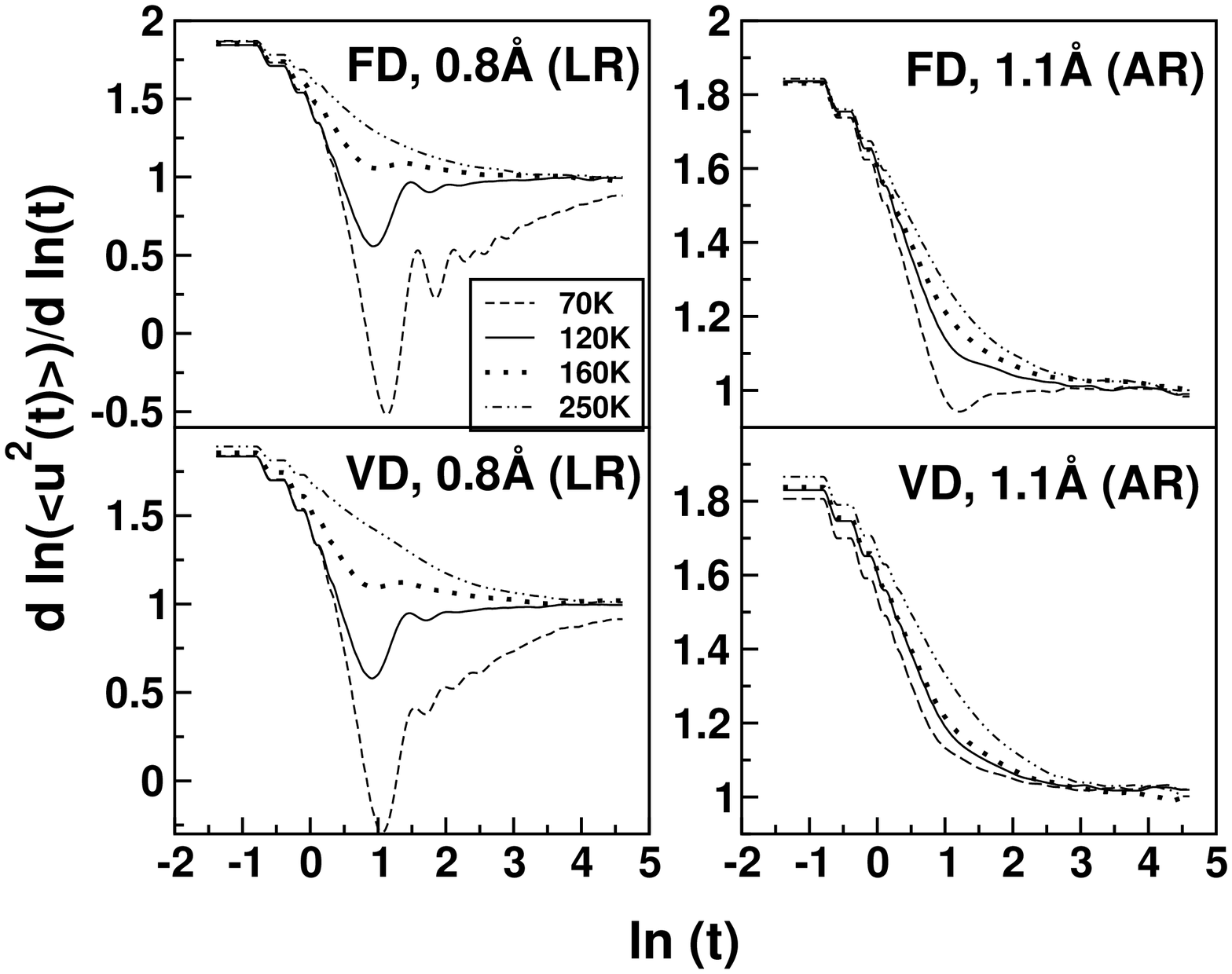}}
\caption {Derivative $d(ln\ u^2(t))/d(ln t)$ as a function of $ln\ t$ ($ln\ t$ = $log_e\ t$)
for solutes belonging to linear(LR) and anomalous(AR) regimes at different temperatures
for the fixed(FD) and varied(VD) density simulations. A comparison of different sizes of 
solutes at a given temperatures for FD and VD sets are shown in (a-b) while
 (c) shows same size at different temperatures for a few sizes.}
\label{deriv_ln_msd}
\end{center}
\end{figure}

In Figure \ref{ln_msd} we show a  plot of $\ln{\langle u^2(t) \rangle}$ against $\ln{t}$ for 
few solute sizes from LR and AR at different temperatures both from FD and VD simulations. 
The following interesting features may be seen from this. There is a transition from ballistic 
to diffusive regime. More importantly, the solute which are from AR (e.g. for sizes 1.1\AA\ and 1.5\AA)
exhibit a rather sharp transition from ballistic to diffusive extending just from $ln\ 0$ to $ln\ 1$. 
The transition region is quite narrow. In contrast, solutes from 
LR exhibit a rather broad ballistic-diffusive transition extending over several picoseconds between
$ln\ 0$ to $ln\ 4$ or higher. This is evident, for example, at 70K (VD) for a size of 0.5\AA. 
For solutes from LR, there is effectively no change in $u^2(t)$ (see Figure \ref{msd}) over a 
significant range of times in the ballistic-diffusive transition regime. Thus, for size 0.5\AA\, 
$0<\ln{t}<1$, the y-axis ($\ln{u^2(t)}$) changes from 1.41 to 1.62 ! This means
that from 1ps to 2.71ps the mean square displacement alters from 4.1\AA$^2$ to 5.1\AA$^2$. The 1.4\AA\ 
sized solute, during the same interval of time has a net change in the mean square displacement of
10.1\AA$^2$ (from 4.3\AA$^2$ to 14.4\AA$^2$). Thus, the solute from LR, is effectively and completely localized 
without any mobility whatsoever during the ballistic to diffusive transition. 

More insight into the nature of ballistic-diffusive transition can be obtained from
the plot of derivative $d(ln\ u^2(t))/d(ln\ t)$ shown in Figure \ref{deriv_ln_msd}. 
The broad ballistic to diffusive transition for solutes from LR and the rather sharp transition
for AR is evident here more clearly than in the plot of MSD.
The changes in the exponent $\alpha$ in ($\langle u^2(t)\rangle \propto t^\alpha$) exhibits some
unusual behaviour as is seen from the figure. First we discuss the behaviour in the face-centred
cubic solid phase. $\alpha$ for solutes from AR vary between
1 and 2 at all temperatures. In contrast to this, $\alpha$ varies widely for solutes from LR,
especially at low temperatures. Note that the $\alpha$ varies only in the subdiffusive
regime ( 0 $<$ $\alpha$ $<$ 1) but it also becomes negative ($\alpha < $ 0) for some time at the low temperature
of 70K, during the ballistic-diffusive transition period. This means that $u^2(t)\ \sim\ 1/\sqrt(t)$
for a short duration during the transition. That is, mean square displacement decreases with
increase in $t$ on the average ! In the liquid phase at 160K, although the exponent $\alpha$ for the 
LR solute varies only between 1 and 2, a bi-exponential decay with $ln\ t$ is seen. For solute from
AR, we see a single exponential decay. At higher temperature (250K), both LR and AR appear to exhibit
a smooth single-exponential decay of the mean square displacement.

We propose here a tentative explanation
why this is may be the case : the forces on the solute from LR exerted by the medium in which it is
located is large (as we shall see below). The ballistic regime is when the solute for a short period
is not experiencing the forces of the surrounding atoms (here essentially the forces due to the
solvent molecules). The diffusive regime is when forces begin to alter the motion of the solute.
However, the large force exerted by the solvent on a solute from LR essentially immobilizes it for
a few picoseconds (during the transition period) while this is not the case for a solute from AR 
which continues to traverse
and at the same time change to diffusive regime. This means that the LR solute is essentially
trapped for an extended period just after the ballistic regime.


The diffusivity dependence on size shows certain changes with density and temperature and melting.

{\it \bf Solid Phase}: At 70K, there are two diffusivity maxima at 1.1 and 1.5\AA\ 
for fixed(FD, $\rho$*=0.933) density system whereas for the variable(VD, 
$\rho$*=1.0803) density system the diffusivity maxima are seen at 1.1 and 
1.4\AA. At a higher temperature of 120K, the  
diffusivity increases from 0.9\AA\ onwards and the second diffusivity 
maximum disappears for the fixed density system(1.5\AA). The second
diffusivity maximum(1.4\AA) exists for variable(VD, $\rho$*=1.0175) density 
system but has lower height as compared to the system at 70K(VD, $\rho$*=1.0803). 

{\it \bf Liquid Phase}: In dense fluid phase at 160K, 
only one diffusivity maximum is observed at  1.1\AA\ for fixed(FD, $\rho$*=0.933) density system 
and variable(VD, $\rho$*=0.8484) density system as compared to two peaks in
solid phase at 70K. The  height of diffusity maximum in liquid phase is
lower than in the solid phase. At 250K, for fixed(FD, $\rho$*=0.933) density,
diffusivity maximum shifts to a lower size of 1.0\AA\ and has a lower 
intensity compared to the system at 160K.
In case of variable(VD, $\rho$*=0.6979) density system at 250K there is a 
gradual decreae in diffusivity with increase in solute size and no 
enhancement in diffusivity for any of the sizes is seen. 

The
anomalous diffusion occurs when the diameter of diffusant is comparable to the 
bottleneck region it passes through during diffusion. 

 A dimensionless parameter
 called levitation parameter, $\gamma$ is defined
 
 \begin{equation}
 \gamma = \frac{\sigma_{opt}}{\sigma_{w}}
 \label{gamma_defn}
 \end{equation}

 \noindent
 where $\sigma_{opt}$ is the diffusant size at which diffusant-medium 
 interactions are optimum and $\sigma_{w}$ is the size of the bottleneck 
 through which the solute passes through during its passage from one void to another. 
In case of systems with a distribution of neck radii, the denominator is replaced
by the average neck radii, $\overline{r}_n$. This is an approximation and does alter
the value of $\gamma$ at which the maximum is seen. Note that existence of a distribution
instead of a well defined single value for the bottleneck radii implies that the 
system is disordered. In the present study, the face-centred solid is relatively more
ordered than the fluid phase but it is still disordered due to thermal motion of atoms
around their equilibrium position. Typically, diffusivity maximum is seen at $\gamma\ \sim\ $1
for ordered solids (without disorder due to thermal motion modeled as rigid framework) 
such as zeolites but at around 0.68 for dense fluids which posses dynamical disorder.\cite{yashosanti94b,
pradipliq}

\begin{figure}
\begin{center}
{\includegraphics*[width=8cm]{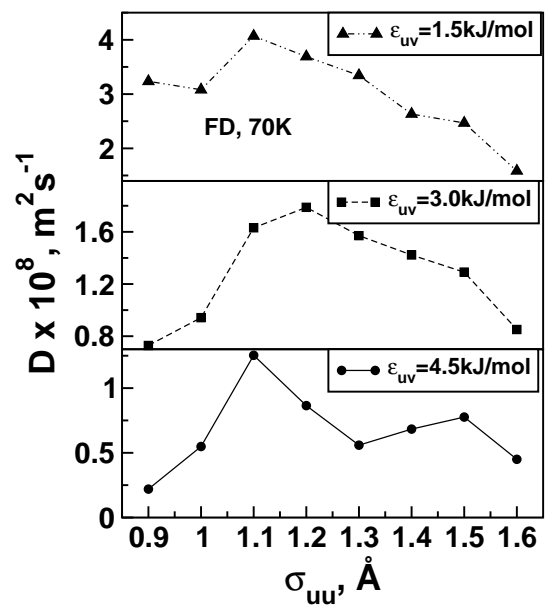}}
\caption{Impurity diffusivity as a function of size for solvent at 70K and
$\rho*$=0.933 (fixed density, FD simulations) and different Lennard-Jones 
solute-solvent interaction parameter, $\epsilon_{uv}$.} 
\label{eps_D}
\end{center}
\end{figure}

Previous studies have shown the importance of diffusant-medium interaction. Absence of
diffusant-medium dispersion interaction leads to complete absence of diffusivity 
maximum. Further, the ratio of $U_{uv}$/$k_BT$ plays an important role. When this ratio
is large, then size dependent diffusivity maximum is seen but when temperature is high
then diffusivity maximum is weak or absent. But, so far, only a single diffusivity maximum
has been seen. However, this is the first study to report existence of double maximum.
We therefore investigated this more carefully. In Figure \ref{eps_D} we
show the results for simulations at 70K for three different values of $\epsilon_{uv}$ = 
1.5, 3.0 and 4.5 kJ/mol. These have been carried out on the smaller sized system
discussed earlier. We see that although a single diffusivity maximum is seen at low values
of diffusant-medium interaction strength, double maximum is seen at large values.
The reasons for this is not clear and requires more detailed study. It is the 
maximum at larger size of 1.5\AA\ which disappears at smaller $\epsilon_{uv}$.

Changes in $D-\sigma$ behaviour as a function of melting may be understood as follows.
We see two maxima at low temperatures and high densities. For example,
the run at 120K (FD) which is at lower density than the VD run exhibits a single maximum. 
These arise from the underlying structure of the solvent as seen in terms of the
void and neck distribution. We see that whenever the distribution of neck sizes $f(r_n)$
is bimodal there are two diffusivity maxima and when $f(r_n)$ shows a single maximum one
diffusivity maximum is seen. However, there are other phases including liquid phase
which show two diffusivity maxima even when $f(r_n)$ is not bimodal \cite{bcc_sy_manju}. 
This is seen whenever $\epsilon_{uv}$ is larger. Clearly, both $f(r_n)$ and solute-solvent interactions
play an important role in giving rise to diffusivity maxima.

At 70K for VD set($\rho^*$ = 1.0803), the bimodal distribution has maxima 
at 0.6 and 0.73\AA\ leading to two maxima in $D$ at 1.1 and 1.4\AA. For FD set which is at lower
density ($\rho^*$ = 0.933), the maxima are at slightly larger size, 0.76 and 0.87\AA\ and 
the maxima in $D$ are at 1.1 and 1.5\AA. The reason why the maximum near 1.1\AA\ did not shift by 0.1\AA\
to larger size could be due to increased disorder at lower density. Effect of such disorder is
to shift the maximum to smaller sizes. Thus, the two opposing effects (change in density and disorder)
lead to no shift. Larger neck size which gives rise to maximum at 1.4\AA\ is relatively less influenced by
the disorder since the neck is quite large. At 120K, for VD ($\rho$*=1.0175) the \frn is still
bimodal with \rn = 0.68 and 0.79\AA and the diffusivity maximum remains at the same position but
height is decreased as compared to 70K. For FD at 120K, we see \frn is no more bimodal and
$D-\sigma$ also exhibits a single maximum. 

In the liquid phase at 160K for variable(VD, $\rho$*=0.8484) density 
system, diffusivity maximum is lower in intensity and occurs at 1.1\AA. 
In this system, average neck radius is 1.01\AA\ which is larger than the 
solid phase systems due to the lower density. 
The fixed(FD, $\rho$=0.933) density system at 
160K has average neck radius of 0.90 and shows anomalous behaviour 
for 1.1\AA. The fact that the maximum in both VD and FD occur at the same value of solute diameter
in spite of the difference in the \rn shows the predominant influence of disorder on
the diffusivity maximum. At 160K, considerable disorder exists in the fluid phase and 
the influence of density is reduced. 

At 250K, variable(VD, $\rho$*=0.6979) density system has maximum 
disorder and in this system solute-solvent interaction energy is not very
large as compared to the kinetic energy($U_{uv}$/$k_BT$ $<<$ 1). Thus, the Levitation effect or
the anomalous diffusion disappears and a gradual decrease in diffusivity with
increase in solute size is observed. But the fixed(FD, $\rho$*=0.933) 
density system at 250K still has density higher than variable(VD) density 
system and this leads to $U_{uv}/k_BT$ is not as small as in the VD run.
Thus,  a small diffusivity maximum is seen for size 1.0\AA.  There is
a shift of the diffusivity maximum to lower size as the system undergoes considerable
decrease in density. These have been summarized in Table \ref{rn_Dmax} which lists
the average values of \rn and the values of $\sigma_{uu}^{max}$ at which the diffusivity maximum
is observed. Also listed are the values of $\gamma$. 

\begin{figure}
\begin{center}
{\includegraphics*[width=6cm]{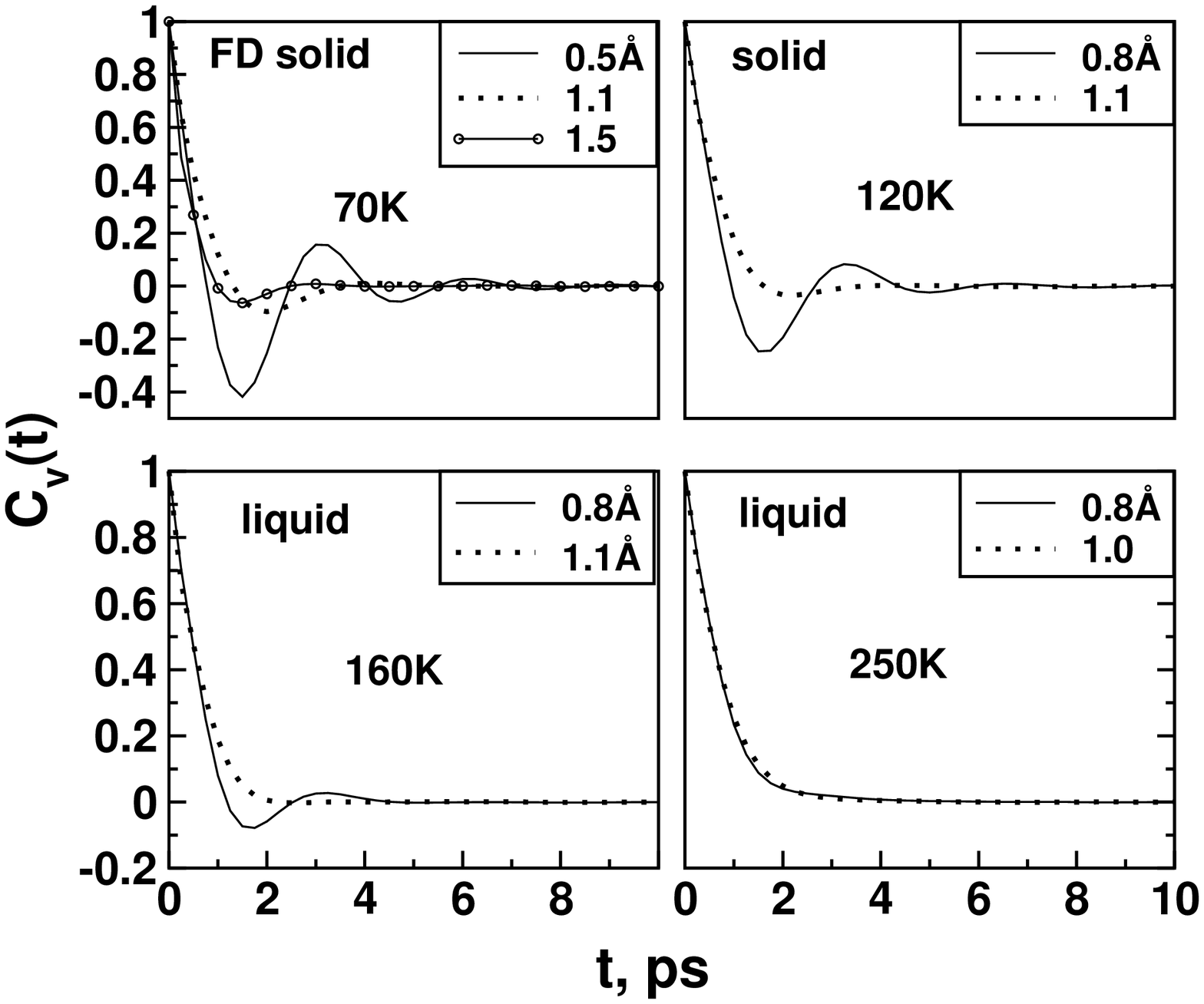}}\hspace*{0.5cm}
{\includegraphics*[width=6cm]{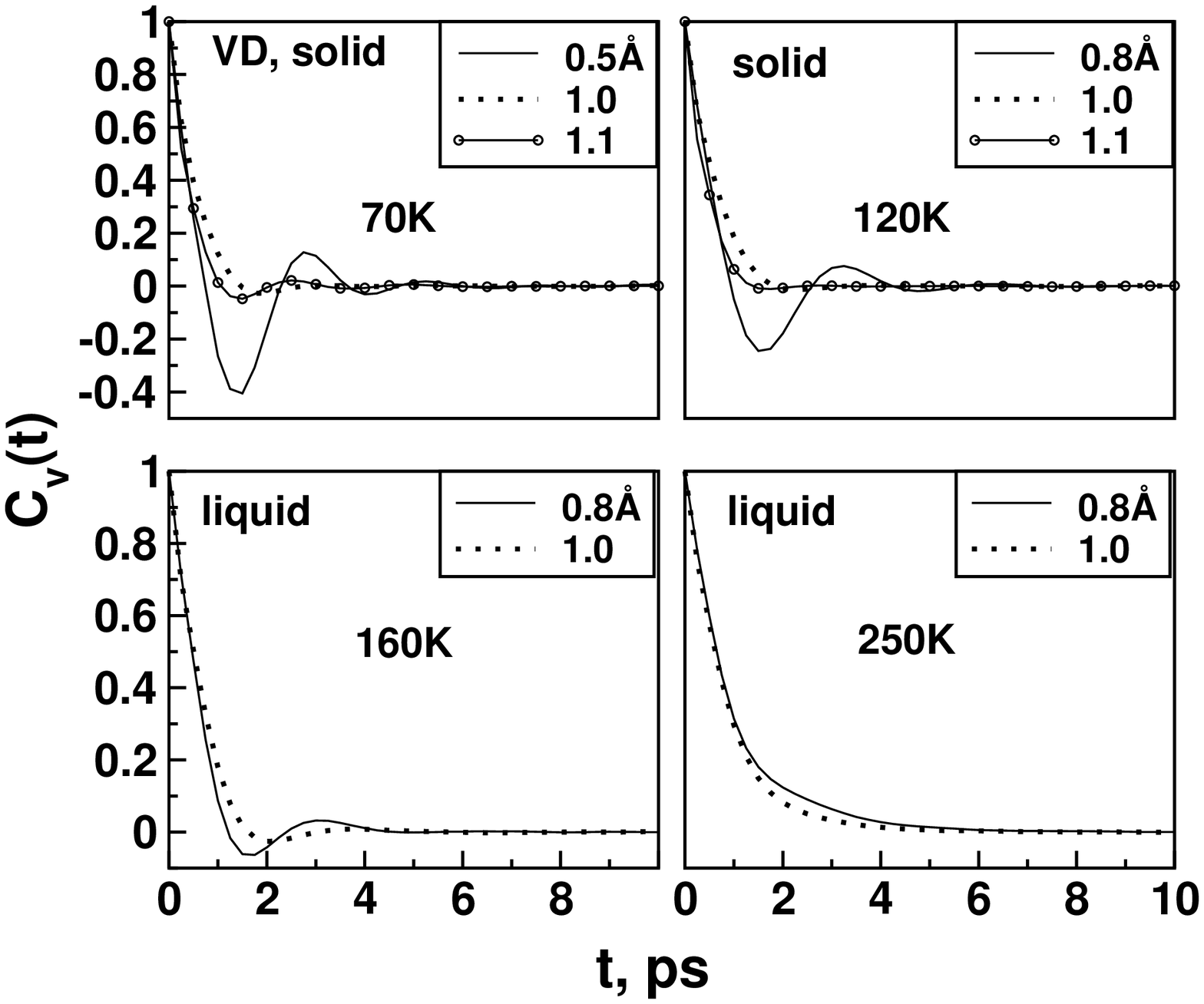}}
\caption{Velocity autocorrelation function of impurities with minimum and maximum
difffusivity. The plots are reported for systems in solid and liquid phases for
fixed and variable densities.} 
\label{vacf}
\end{center}
\end{figure}

We have obtained the related properties to obtain additional insight. 
Figure \ref{vacf} displays the velocity autocorrelation(VACF)
function for solutes in all the systems. Solutes corresponding to minimum 
diffusivity, (0.5\AA\ for FD and VD at 70K) and 0.8\AA\ for other systems 
show an oscillatory VACF with significant back scattering while solutes at 
diffusivity maximum decay smoothly. We attribute oscillatory behaviour for solutes from
LR to an energy barrier during its passage out of the first solvent shell.
Such a barrier leads to back-and-forth motion leading to back scattering.
The anomalous regime solute does not experience any barrier at the first shell
and consequently, VACF decays smoothly. In the solid
phase, a slight back scattering is seen for the solute from AR. This could be either due to the
discrete sizes of solutes (at 0.1\AA\ interval) for which properties have been
computed here. The actual diffusivity maximum might be actually somewhere in between
these. Alternately, the cooperative motion in a liquid between the solute and solvent is
less likely in a solid since the solvent particles in the solid phase have limited mobility.

\begin{figure}
\begin{center}
{\includegraphics*[width=6cm]{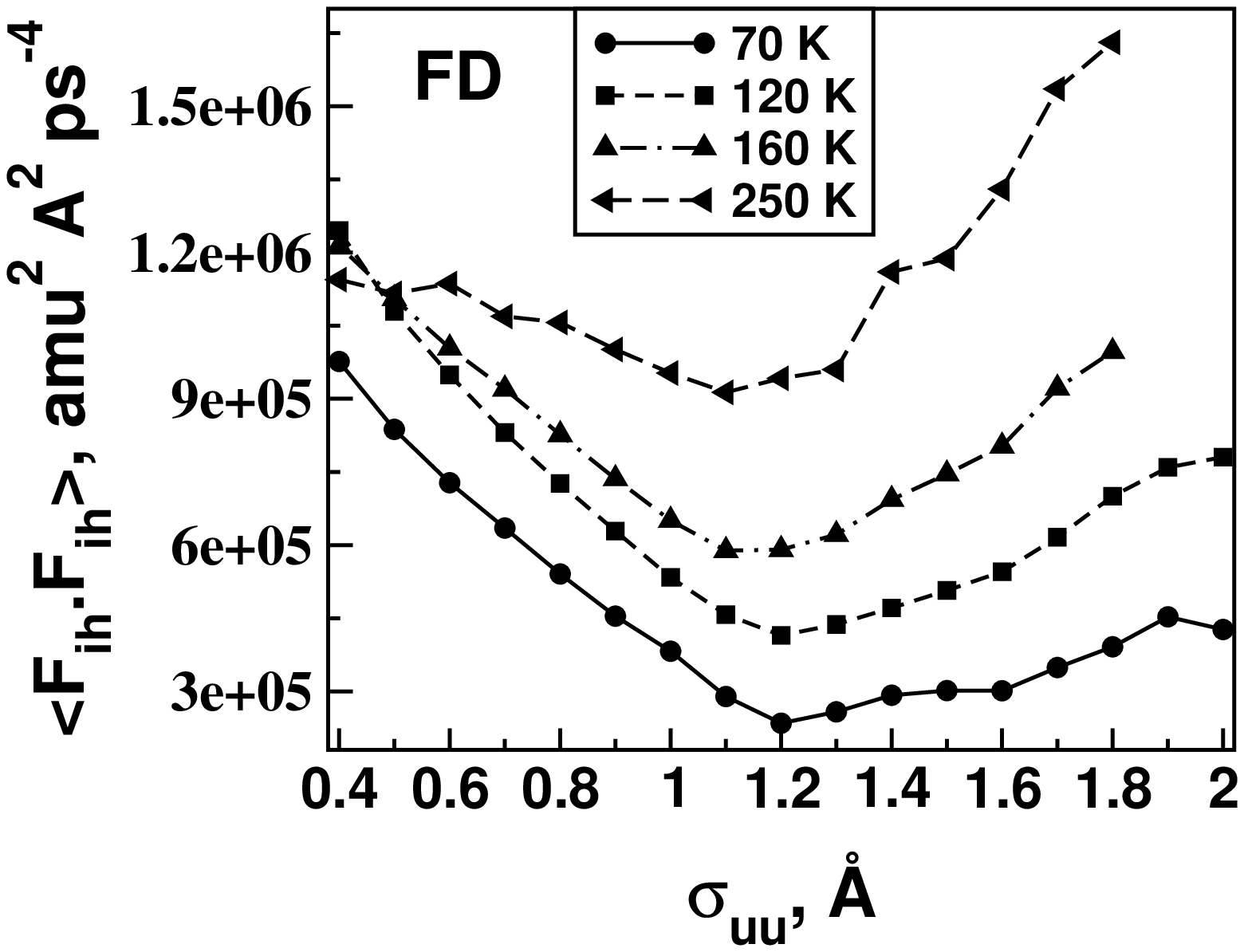}}\hspace*{0.5cm}
{\includegraphics*[width=6cm]{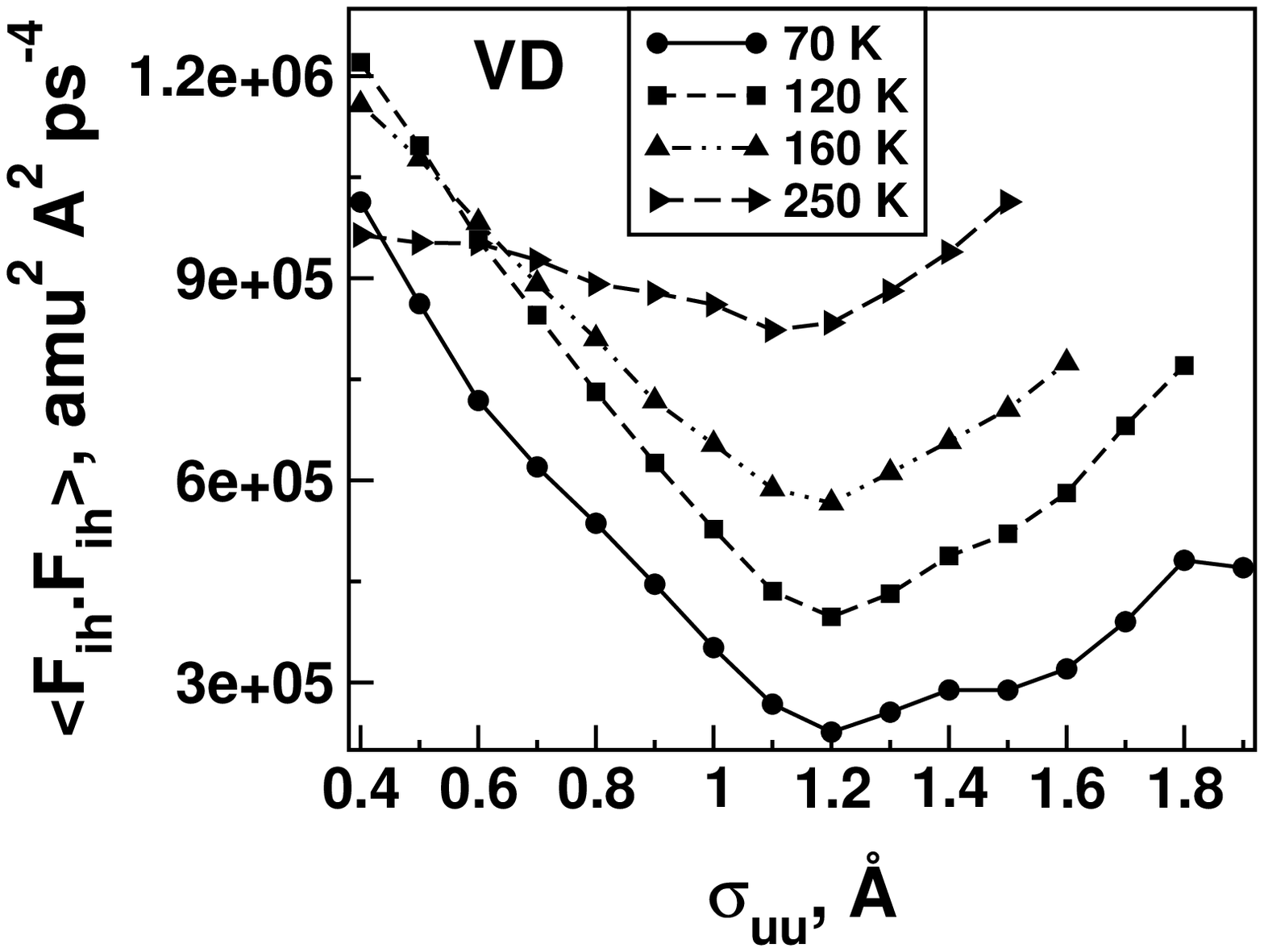}}
\caption{ Average mean square force occuring on an solute due to 
neighbouring solvent atoms as a function of solute size for systems in 
different phases for fixed and variable densities.}
\label{avfsq}
\end{center}
\end{figure}

Figure \ref{avfsq} plots the average mean square force exerted by the surrounding
atoms (predominantly solvent particles) as a function of the solute
size. Solutes with maximum diffusivity experience lowest average mean square
force  due to the neighboring solvent atoms in all the systems. 
When the size of the solute is similar to the bottleneck through which it 
passes during diffusion, the force from a given direction is equal and opposite to that
exerted by the solvent atoms diagonally opposite. The force from diagonally opposite directions
mutually cancel each other and solute  has a higher diffusivity. Note that the average
mean square force is lowest (for all sizes of the solute) in the solid phase at 70K and with increase 
in temperature, there is increase in net force. This is due to increasing disorder or Brownian
motion which undermines the intermolecular forces. Random forces form a significant fraction of
total forces at high temperature.  As the difference in the force exerted between the solute from
LR and AR decreases, the diffusivity maximum also gradually disappears.

\begin{figure}
\begin{center}
{\includegraphics*[width=6cm]{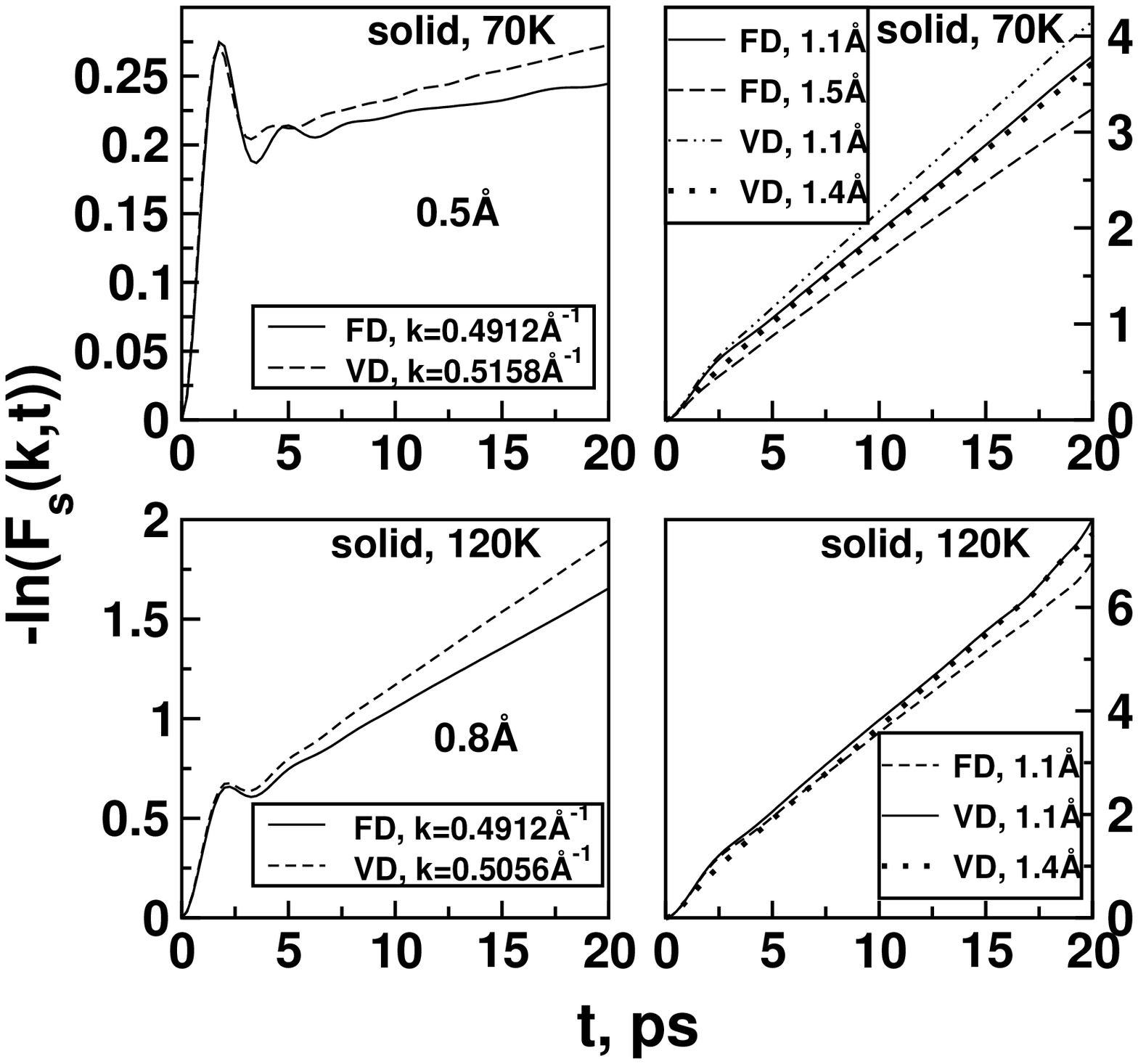}}\hspace*{0.5cm}
{\includegraphics*[width=6cm]{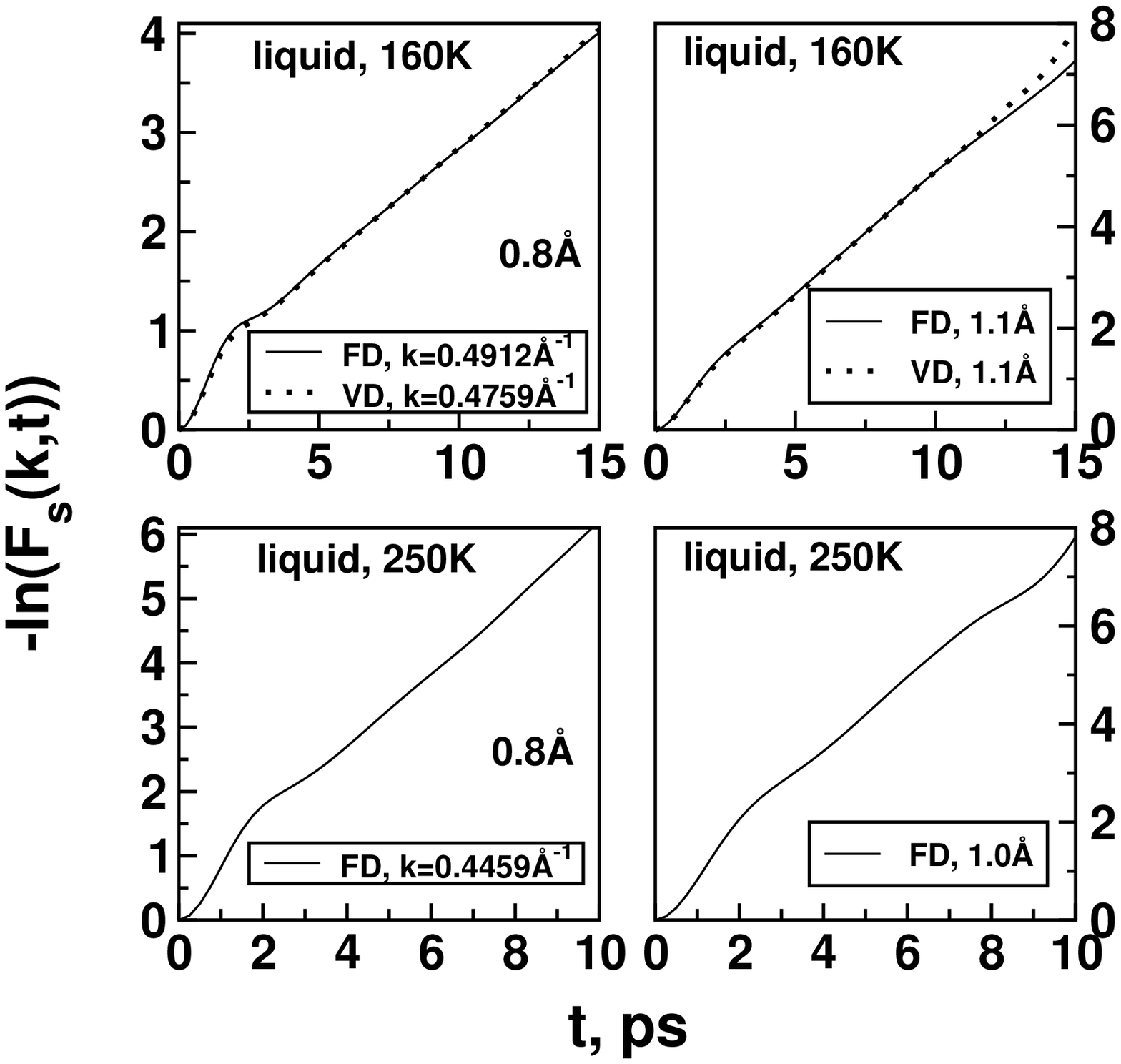}}
\caption{Logarithm of self part of the intermediate scattering function 
is shown for wavenumber $k$ = 2n$\pi$/L for solute sizes corresponding to
minimum and maximum diffusivity in both the solid and liquid phase obtained from
variable and fixed density runs.  
L($\rho^*$)=51.1725\AA(0.933, FD), 48.7327\AA(1.0803, VD(70K)), 
49.7153\AA(1.0175, VD(120K)), 52.819\AA(0.8484, VD(160K)) and 
56.37\AA(0.6979, VD(250K)). The lines are the least square fit to the MD
data.}
\label{log_fskt}
\end{center}
\end{figure}

Time evolution of the negative of logarithm of self part of the intermediate scattering 
 functions $F_s(k,t)$ for solute sizes at diffusivity maximum and close to diffusivity minimum 
at the border between the LR and AR are shown in Figure \ref{log_fskt}. 
The reciprocal of the slope gives the relaxation time.\cite{jacucci_rahman}
Note that solutes away from the diffusivity maximum exhibit two distinct slopes; 
at initial times the slope is
large and at long times a lower slope corresponding to two relaxation times.
In contrast, solute close to diffusivity maximum exhibits either a single slope
or two straight lines whose slopes are almost identical to each other. 
The biexponential decay of $F_s(k,t)$ for solutes from LR suggest that the motion of
the solute has two distinct regime. We suggest that the faster relaxation time
of this solute corresponds to motion within the first neighbour shell of the solvent
irrespective of whether the phase is solid or fluid. For solute from AR, we suggest that the 
solvent motion encounters no barrier past the first solvent shell and therefore there is
a single relaxation time. The resulting relaxation times are listed in Table \ref{relax_time}. With increase
in temperature, the relaxation times decrease due to faster dynamics.

\begin{table}
\caption {Relaxation times of impurities belonging to linear and anomalous
regimes in different systems.}
\begin{center}
\begin{tabular}{{c}{c}{c}{c}{c}{c}{c}}\hline
&\multicolumn{3}{|c}{Fixed density}&\multicolumn{3}{|c}{Variable density}\\ \hline 
T, K & $\sigma_{uu}$,\AA & k, \AA$^{-1}$ & $\tau_1$, $\tau_2$ (ps)&$\sigma_{uu}$,\AA & k, \AA$^{-1}$ & $\tau_1$, $\tau_2$ (ps) \\\hline\hline
70 & 0.5 & 0.4912 & 5.68, 409.84 & 0.5 & 0.5158 & 5.69, 263.16 \\
70 & 1.1 & 0.4912 & 5.42 & 1.1 &0.5158 & 4.92 \\
70 & 1.5 & 0.4912 & 6.19 & 1.4 & 0.5158 & 5.45 \\
120 & 0.8 & 0.4912 & 2.59, 16.53 & 0.8 & 0.5056 & 2.50, 13.61 \\
120 & 1.1 & 0.4912 & 3.06 & 1.1 & 0.5056 & 2.77 \\
120 &     &       &      & 1.4 & 0.5056 & 2.74 \\
160 & 0.8 & 0.4912 & 1.72, 4.25 & 0.8 & 0.4759 & 1.83, 4.18 \\
160 & 1.1 & 0.4912 & 2.10 & 1.0 & 0.4759 & 1.99 \\
250 & 0.8 & 0.4912 & 1.02, 1.77 & & & \\
250 & 1.0 & 0.4912 & 1.33 & & & \\

\end{tabular}
\label{relax_time}
\end{center}
\end{table}

\begin{figure}
\begin{center}
{\includegraphics*[width=6cm]{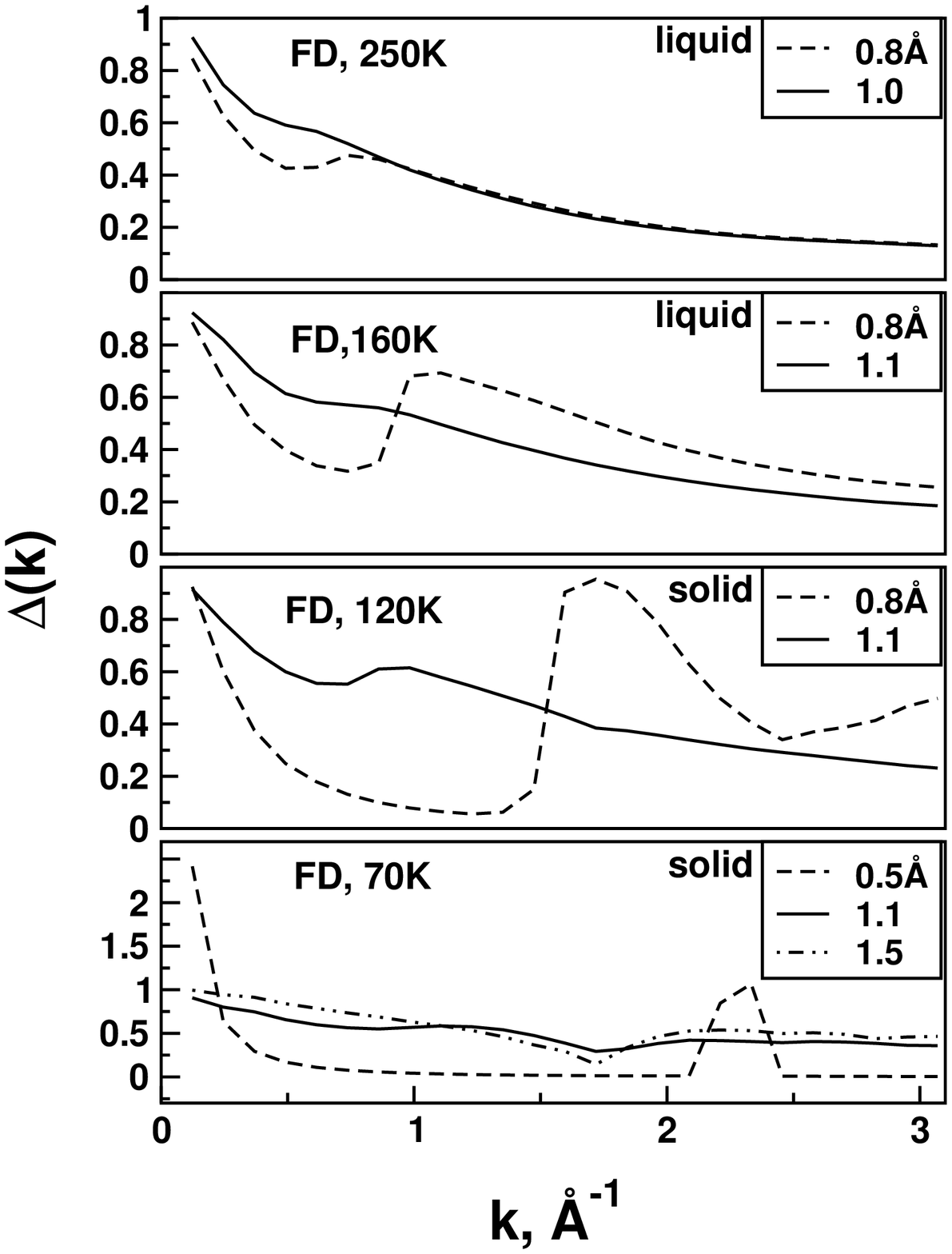}}\hspace*{0.5cm}
{\includegraphics*[width=6cm]{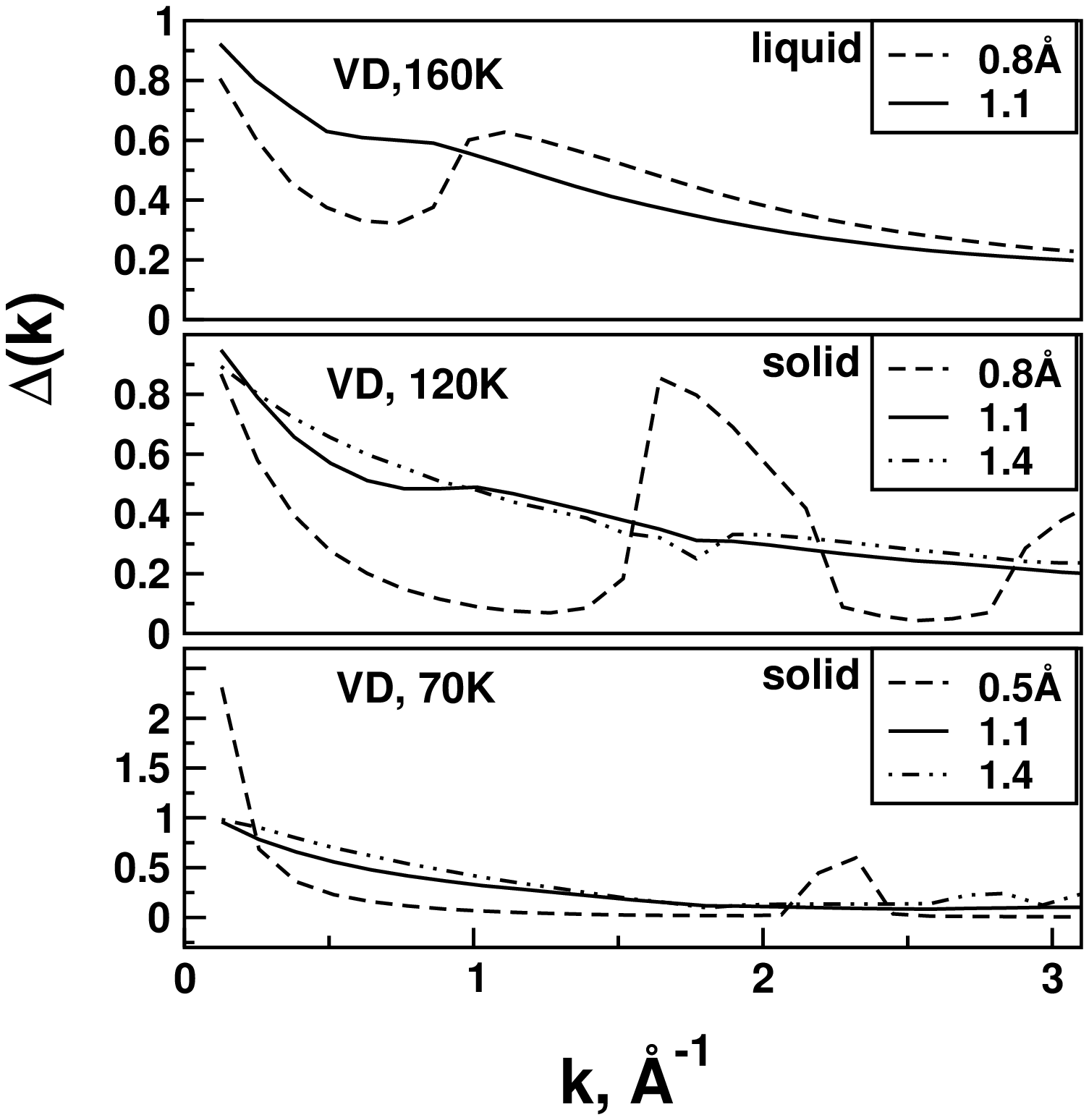}}
\caption{Wavevector dependence of impurities with minimum and maximum diffusivity 
in solid and liquid phase at fixed and variable density.}
\label{delta_k}
\end{center}
\end{figure}

Full width at half maximum (fwhm)  of the dynamic structure factor has been obtained by 
fourier transformation of the self  part of the intermediate scattering 
function. The dependence of fwhm on $k$ gives insight into the nature of $k$-dependence of
self diffusivity. In the limit of $k$ approaching 0, the hydrodynamic limit,
the fwhm approaches 2D$k^2$ and the ratio 

\begin{equation}
\Delta(k) = \frac{\Delta\omega(k)}{2Dk^2}
\label{deltaomega_eq}
\end{equation}

approaches unity. Figure \ref{delta_k} shows the wavevector 
dependence of self diffusivity for solutes close to to minimum and maximum diffusivity,
both for solid and liquid phases. $\Delta$(k) of solute close to diffusivity maximum
exhibit a monotonic decrease with $k$ both in the solid  and the liquid phase.
In contrast, the solute near the diffusivity minimum exhibit a minimum followed by
a maximum for $k$ between 1 and 3. The minimum corresponds to slow down in diffusivity
of the solute at wavenumber corresponding to first shell. Similar behaviour has been
seen by Nijboer and Rahman in high density liquid argon at low temperatures \cite{rahman66}. 
Boon and Yip have noted that the minimum generally occurs for wavenumbers 
corresponding to first maximum in static structure factor S(k) \cite{boon_yip}. Such a slowing down
in $k$ dependent self diffusivity does not occur for solute from AR since it does
not encounter any energetic barrier to go past the first solvent shell. Consequently,
it has facile passage past the first shell. If the solute size is close to the maximum
in diffusivity but not at the {\em true} maximum (the location of the true maximum is limited by
the 0.1\AA\ interval we have used here for the solute size) then small oscillations
are seen in Figure \ref{delta_k}. 

\begin{figure}
\begin{center}
{\includegraphics*[width=8cm]{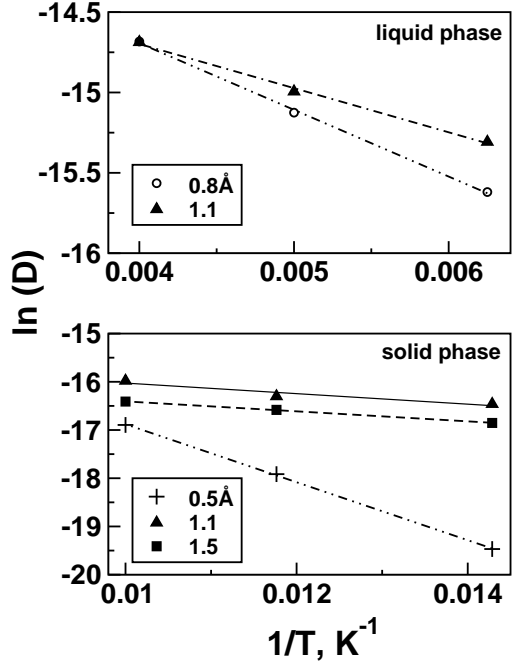}}
\caption{ Arrhenius plot for impurities with diffusivity minimum and maximum 
in solid and liquid phase for fixed density systems.}
\label{arrh}
\end{center}
\end{figure}

We have obtained the activation energies of the solute near diffusivity
minimum(0.5\AA\ and 0.8\AA) and maximum(1.1\AA\ and 1.5\AA) 
in the solid and liquid phases for the fixed
density systems. Figure \ref{arrh} displays the Arrhenius plots for the  
solutes. Diffusivities of solutes have been computed from additional simulations
at temperatures 70, 85 and 100K in the solid phase and 160, 200 and 250K in the liquid phase. The 
activation energy for diffusion is less for solutes with maximum diffusivity
as compared to solutes located at minimum diffusivity as shown in Table \ref{Eact}. 
The difference in activation energy for minimum and maximum diffusivity is
very large in the solid phase. In the liquid phase, system is
dynamic and overall diffusivity of impurities of all sizes is higher than the
solid phase. However, the increased influence of random forces leads to 
limited influence of interparticle interactions on dynamics. Thus, the difference in
activation energies between the solute from LR and AR decrease. This is seen
for the liquid phase for sizes 0.8 and 1.1\AA. In case of solid phase, due to the
predominance of solute-solvent interaction and lower kinetic energy, the difference
in activation energies for solutes from LR and AR are dramatic : 4.99 kJ/mol for
0.5\AA\ solute from LR as compared to 0.86 kJ/mol for the {\em larger} 1.5\AA\ sized solute.
This shows that the Levitation effect is more pronounced in the solid phase.

\begin{table}
\caption {Activation Energy of impurities with diffusivity minimum and maximum 
in solid and liquid phases for fixed density systems.}
\begin{center}
\begin{tabular}{{c}{c}{c}{c}}\hline
$\sigma_{uu}$, \AA &Phase&Regime &$E_{act}$, kJ/mol\\\hline\hline
0.5&Solid&Linear&4.9998\\
1.1&Solid&Anomalous&0.9070\\
1.5&Solid&Anomalous&0.8625\\
0.8&Liquid&Linear&3.4525\\
1.1&Liquid&Anomalous&2.2852\\ \hline
\end{tabular}
\label{Eact}
\end{center}
\end{table}

\section{Conclusions}

The main conclusions of this study are : (i) At sufficiently low density size dependent 
diffusivity maximum disappears altogether. (ii) Presence of dynamic disorder at
high temperature does not lead to backscattering in VACF for the solute close to diffusivity maximum.
(iii) Significant backscattering is seen in VACF for LR but not for AR in the solid phase.
(iv) In the solid phase, the exponent $\alpha$ in $u^2(t)\ \sim\ t^\alpha$ exhibits
negative exponent during ballistic to diffusive transition for the solute from LR but not from AR.
(v) The solute diameter $\sigma_{uu}^{max}$ (equal to $\sigma_{uu}$ for the solute at diffusivity
maximum) shifts to lower sizes at high temperature in spite of decrease in density. This is 
because of increase in disorder. (vi) Multiple maxima are seen in the solid phase
for high $\epsilon_{uv}$. (vii) Significant shift to lower \rn is seen across the solid-liquid transition
corresponding to lower $\rho$. (viii) Correspondigly, shift is seen in $\sigma_{max}$ to smaller
values. 

\noindent
{\em Acknowledgement} : Authors wish to thank Department of Science and
Technology, New Delhi for financial support in carrying out this work.
Authors also acknowledge C.S.I.R., New Delhi for a research fellowship to
M.S.

\bibliographystyle{plain}
\bibliography{manju_PT}

\end{document}